\documentclass[letterpaper,tight,classic,final,pageclassic,orcish,nobm]{iopart}
\usepackage{graphicx,graphicx,epsfig} 

\begin{document}

%\newif\ifpdf 
%\ifx\pdfoutput\undefined
%\pdffalse % we are not running PDFLaTeX
%\else
%\pdfoutput=1 % we are running PDFLaTeX
%\pdfcompresslevel=9
%\pdftrue
%\fi

%\documentclass[10pt,a4paper,twocolumn,multicol,superscriptaddress,prb,amsmath,amssymb,aps,prb]{revtex4}
%\documentclass[10pt,a4paper,superscriptaddress,prb,amsmath,amssymb,aps,prb]{revtex4}
%\usepackage{axodraw}
%\usepackage{graphicx}
%\usepackage{bm}
\newcommand{\rr}{{\bf r}}
\newcommand{\uu}{{\bf u}}
\newcommand{\kk}{{\bf k}}
\newcommand{\RR}{{\bf R}}
\newcommand{\qq}{{\bf q}}
\newcommand{\QQ}{{\bf Q}}
\newcommand{\GG}{{\bf G}}
\newcommand{\dd}{{\vec{\bf \delta}}}
\newcommand{\cc}{{\hat c}}
\newcommand{\ah}{{\hat a}}
\newcommand{\bb}{{\hat b}}
\newcommand{\znpk}{z_{n'k}}
\newcommand{\znk}{z_{nk}}
\newcommand{\ccd}{{\hat c^\dagger}}
\newcommand{\ahd}{{\hat a^\dagger}}
\newcommand{\bbd}{{\hat b^\dagger}}
\newcommand{\la}{\langle}
\newcommand{\ra}{\rangle}
\newcommand{\up}{\uparrow}
\newcommand{\dn}{\downarrow}
\newcommand{\rar}{\rightarrow}
\def  \bsig    {\mbox{\boldmath$\sigma$}}
               \def  \btau    {\mbox{\boldmath$\tau$}}

%\title
\review{Semiclassical theories of the anomalous Hall effect.}
\author{N.A. Sinitsyn}
%\shortauthor{N.A. Sinitsyn and Ilya Nemenman}

\address{
  Center for Nonlinear Studies and 
  Computer, Computational and Statistical Sciences Division, 
  Los Alamos National Laboratory, Los Alamos, NM 87545 USA
}
%{Phases: geometric; dynamic or topological}
%\pacs{05.10.Gg}%{Stochastic analysis methods (Fokker-Planck, Langevin, etc.)} 
%\pacs{05.40.Ca}%{Noise}

%\author{ N.A. Sinitsyn and I. Nemenman.}
%\affiliation{Center for Nonlinear Studies and \\
%Comuputer, Computational and Statistical Science Division\\
%Los Alamos National Laboratory, Los Alamos, NM 84545}
%\date{today}

\begin{abstract}{
Recently, the semiclassical theory of the anomalous Hall effect induced by the Berry curvature in Bloch bands
has been introduced. The theory operates only
with gauge invariant concepts, that have a simple semiclassical interpretation and
provides a clear distinction among various contributions to the  
Hall current.
While the construction of such an approach to the anomalous Hall effect problem
has been long sought, only the new semiclassical theory demonstrated the agreement 
with quantitative results of rigorous approaches based on the Green function
techniques. The purpose of this work is to review the semiclassical approach including the early 
ideas and the recent achievements.}
\end{abstract}
\tableofcontents 

\maketitle

\section{Introduction}
The anomalous Hall effect (AHE) is one of the most famous transport phenomena in
magnetic materials. Unlike in paramagnets, the Hall resistance $R_{xy}$ of a magnetic film has two 
contributions. One is usual, it is proportional to the applied magnetic field $H$. 
The other one is anomalous, 
it is observable only in a ferromagnetic state. The anomalous contribution is often proportional to the
magnetization rather than to the applied magnetic field
\begin{equation} 
R_{xy}=r_0H+r_aM,
\label{rxy}
\end{equation}
where $M$ is the magnetization of the sample, $r_0$ and $r_a$ are constants that characterize the strength of the standard and
the anomalous Hall resistivities respectively. The recent theoretical research demonstrated that the
 linear dependence on $M$ in (\ref{rxy})  is not universal and the Hall resistivity can show resonance features as a function of variable parameters, including
the magnetization. 

The practical interest in the AHE has continuously been driven by a difficulty to measure the carrier density in ferromagnets. Standard techniques based on 
measurements of the Hall conductivity are obstructed by the considerable anomalous contribution which can be much greater than the
standard Hall conductivity. 
%due to the Lorentz force on the carriers.
 Recent advances in spintronics, especially the 
creation of new types of diluted magnetic semiconductors,  revived the interest in the AHE as a useful tool to control 
spin-polarized currents and to characterize the magnetization. 
In addition, the recent theoretical interest has been fueled by the  new interpretation of the anomalous Hall conductivity 
in terms of Berry phases and topological defects in the crystal band structure. Many theoretical constructions
that usually had been considered of relevance 
mainly in high energy physics such as  
noncommuting coordinates and magnetic monopoles,  became useful and even measurable in experiments on the AHE  \cite{jungwirth-02, fang-03, experiment-ahe1, yao-04}.

Despite the long history and the considerable practical importance, the theory of the anomalous Hall effect has remained controversial.  
The first steps to explain the AHE in ferromagents were made more than 50 years ago. Since then many articles were published to
 correct previous mistakes and to
suggest new explanations. Many of such efforts still were incomplete. While they resolved several pieces of the puzzle they also
disregarded the others. Sometimes distinct quantitative predictions followed from applications of different  methods to the same model. Such a 
controversy persists even at the present time. For example, there is a number of recent publications with contradictory quantitative predictions   
for the AHE in the Rashba 2D electron system
 \cite{culcer-03,dugaev-05,sinitsyn-05,liu-05,liu-06,inoue-06,onoda-06,borunda-07,nunner-07,kato-07};
although the issue has finally been resolved \cite{nunner-07,kato-07}.

Many aspects of the AHE have been extensively reviewed in the literature.
 The detailed up to date discussion of experiments with diluted magnetic semiconductors and the 
comparison with existing numerical and theoretical predictions can be found in recent reviews \cite{dietl-rev,sinova-rev1, sinova-rev2, fukumura-rev}.
The modern topological interpretation of the AHE and the Kubo formula in terms of a magnetic monopole  in the momentum space
was reviewed in \cite{nagaosa-rev}.  There are also much older introductions \cite{hurd-rev,berger-rev}, concentrated on the side-jump effect in III-V
 semiconductors,
 although many
concepts, discussed there, have been strongly revised in recent years. The recent review of these older work, made in the same spirit, can also be found in  \cite{engel-rev} together with 
the discussion of results on the 
spin Hall effect.

Numerous efforts to
design a rigorous semiclassical approach that would explain in simple terms all possible contributions to the anomalous Hall conductivity including disorder effects, however, still
lack a detailed comparison in a single work. 
 The goal of the present review is to fill this gap and to discuss in more details
the existing semiclassical theories, their advantages and limitations. 
We start with the earliest ideas introduced by Karplus, Luttinger and Smit and end with the most recent constructions that demonstrated the 1-1 agreement with the  rigorous quantum mechanical
techniques.

The structure of this review is the following. In the rest of the introduction in section 1.1, we recall the basic text book information about the semiclassical approach to conductivity 
calculations taking as an example free electrons interacting with elastic scatterers. In section 1.2,  I remind several commonly known facts about Bloch bands and explain how the introduction of the
band structure complicates the creation of the semiclassical theory of the transport, especially in the application to the AHE.
 In section 2, I discuss the forces driving the AHE and review the 
earliest theories, including  the quantum analog of the Boltzmann equation applied by Luttinger (section 2.1) and the introduction of noncommuting coordinates by Adams and Blount (section 2.2).
In section 3, I proceed with more recent theories based on the gauge invariant formulation of the wave packet dynamics and its Berry phase interpretation. In sections 3.1
and 3.2,
I will review the application of this approach to the anomalous Nernst effect, the intrinsic contribution contribution and the side-jump effect.
Sections 4 and 5 are devoted to the rigorous semiclassical theory of the AHE, free of most limitations of previous approaches. In section 4, I
discuss the rules that connect the scattering 
matrix with the classical concepts such as the scattering probability and the size of the
coordinate shift at a scattering event.  In section 5, I introduce the semiclassical Boltzmann equation and subsequently explain
all important contributions to the Hall conductivity. There I will also discuss the strength of the AHE and comparisons
with rigorous quantum mechanical approaches. Section 6 is the summary that discusses the present status of the theory and outlines possible future research directions. 
 
\subsection{Semiclassical approach to conductivity calculations.}  
% The semiclassical approach is
%based on the understanding 
%the dynamics of wave packets in the phase space. Their behavior sometimes can be explained by purely classical means.
Quantitative estimates of  the dc AHE by standard techniques based on the evaluation of Green functions and their products invariably involve 
long complex calculations. It is hard to achieve  
 transparent interpretations; therefore
theories of the AHE normally focus on particular simple model Hamiltonians
and ignore many-body interactions apart from mean-field exchange potentials that encode the magnetic order.
Even with these simplifications,  the AHE  theory remains 
difficult to develop.

One of the problems is  the small magnitude of the AHE in comparison to the longitudinal conductivity. When
considering the perturbative expansion of the conductivity in the weak disorder limit, the AHE contribution appears only in subdominant terms
 of higher powers in small parameters. Many standard approximations turn out to be no longer valid at these orders. Even a proper counting of  relevant
terms of a similar strength was a problem in many cases.
 Another difficulty is in the physical interpretation of conductivity contributions in the Kubo formula or in the Keldysh technique. 
Generally these rigorous quantum mechanical approaches operate with nongauge-invariant objects, such as  off-diagonal elements of  Green functions, of the density matrix or of the velocity
operator, which
only in the end  are combined in the gauge invariant expression for the conductivity. Such calculations, while formally rigorous, hide the physical origin of elementary
microscopic processes. This complicates the analysis and the bookkeeping of the relevant contributions.

The alternative approach is based on the classical Boltzmann equation applied to the electron transport  \cite{mott-book,ashcroft-book,loss-03}. 
It can be justified by the fact that in sufficiently clean materials one can look at the transport from the basis of wave packets rather than Bloch waves.
In the dilute disorder limit,  a wave packet is not
destroyed during long time and behaves in many respects as a classical particle. One can trace  the motion of the wave packet in external fields and describe it
in terms that have a clear meaning in  classical physics.

In crystals, electrons cannot be considered as free particles because they strongly interact with a periodic crystal potential. As we will see in following 
sections this creates interesting ingredients in  the wave packet dynamics but initially, we will describe the semiclassical theory free of these complications assuming
that electrons do not interact with the crystal potential and with each other \cite{mott-book}.
The impurity free  Hamiltonian of such an electron system has plain wave eigenstates
\begin{equation}
\psi_{{\bf k}}({\bf r},t)=\frac{1}{L^{D/2}}e^{i{\bf k \cdot r}-i\frac{k^2}{2m}t},
\label{plain}
\end{equation}
where $L$ is the size of the system,  $D$ is its spatial dimension and $k=|{\bf k}|$.
To construct a wave packet with a well defined average momentum
${\bf k_c}$, plain waves (\ref{plain}) should be superposed with 
the envelope function $a({\bf k})$, sharply peaked  near the point
${\bf k}={\bf k_c}$ so that $\int d^D{\bf k} |a({\bf k})|^2 {\bf k}  ={\bf k_c}$, then the wave packet vector can be written in the coordinate representation as follows
\begin{equation}
\Psi_{{\bf k_c}}({\bf r},t) = 
\int \frac{d^D{\bf k}}{L^{D/2}} a({\bf k})exp \left\{ i\left( {\bf k \cdot r}-\frac{k^2t}{2m}  \right)\right\}.
\label{wp1}
\end{equation}
The normalization condition requires that 
\begin{equation}
\langle \Psi_{{\bf k_c}}|\Psi_{{\bf k_c}}\rangle= \int d^D{\bf r} \Psi_{{\bf k_c}}^*({\bf r},t) \Psi_{{\bf k_c}}({\bf r},t) =\int d^D{\bf k}
 |a({\bf k})|^2=1
\label{norm}
\end{equation}
and the index ${\bf k_c}$ tells that the wave packet has this average momentum, namely, switching to the momentum representation one can find that

\begin{equation}
\fl
{\bf k_c}= \langle \Psi_{{\bf k_c}}| \hat{{\bf k}} |\Psi_{{\bf k_c}}\rangle=
\int d^D{\bf k} \int d^D{\bf k'}
a({\bf k})a^*({\bf k'}) 
\langle {\bf k'} |\hat{\bf k}   |{\bf k} \rangle =
\int d^D{\bf k} |a({\bf k})|^2 {\bf k} .
\label{mom}
\end{equation}

The velocity of the free wave packet center of mass can be derived as follows
\begin{equation}
\fl
\begin{array}{l}
\dot{{\bf r}}_c= \frac{d}{dt} 
 \langle \Psi_{{\bf k_c}}|  \hat{{\bf r}} |\Psi_{{\bf k_c}}\rangle=
\frac{d}{dt}\left\{ \int \frac{ d^D{\bf r}}{L^{D}} \int d^D{\bf k} \int d^D{\bf k'}
a({\bf k})a^*({\bf k'}) e^{-i{\bf k'r }}\left( {\bf r}e^{i{\bf kr }}\right)e^{i\frac{(k')^2t}{2m}-i\frac{k^2t}{2m}} \right\}  = \\
\\\,\,\,\,\,\,\,\,\,\,\,\,\,\,\,\,\,\,\,\,\,\,\,\,\,\,\,\,\,\,\
\frac{d}{dt}\left\{  \int d^D{\bf k} \int d^D{\bf k'}
a({\bf k})a^*({\bf k'}) \left (\int \frac{ d^D{\bf r}}{L^D} e^{-i{\bf k'r }} \left(-i\frac{\partial}{\partial \bf k}e^{i{\bf kr }}
\right)\right)e^{i\frac{(k')^2t}{2m}-i\frac{k^2t}{2m}} \right\}  = \\
\\\,\,\,\,\,\,\,\,\,\,\,\,\,\,\,\,\,\,\,\,\,\,\,\,\,\,\,\,\,\,\,\,
\frac{d}{dt} \left\{ \int d^D{\bf k} \int d^D{\bf k'}
a^*({\bf k'}) \delta({\bf k} -{\bf k'} ) e^{i\frac{(k')^2t}{2m}}i\frac{\partial}{\partial {\bf k}} [a({\bf k}) e^{-i\frac{k^2t}{2m}}] \right\}=\\
\\\,\,\,\,\,\,\,\,\,\,\,\,\,\,\,\,\,\,\,\,\,\,\,\,\,\,\,\,\,\,\,\,\,
\frac{d}{dt}\left\{ \int d^D{\bf k}
 |a({\bf k})|^2 \frac{{\bf k}t}{m}   \right\}+ \frac{d}{dt}\left\{ \int d^D{\bf k}
 a({\bf k})^* \left( \frac{-i\partial }{\partial  {\bf k} }  a({\bf k}) \right) \right\}  = \frac{{\bf k_c}}{m}.
\end{array}
\label{vel}
\end{equation}

In the external uniform electric field, the Hamiltonian operator is $\hat{H}=\hat{k}^2/2+e{\bf E \cdot }\hat{{\bf r}}$ and the average of the momentum is changing
with time
\begin{equation}
\fl
%\begin{align}
\begin{array}{l}
\dot{{\bf k}}_c= \frac{d}{dt} 
 \langle \Psi_{{\bf k_c}}|  \hat{{\bf k}} |\Psi_{{\bf k_c}}\rangle_{{\bf E}}=\frac{d}{dt} \left\{ \int d^D{\bf k} \int d^D{\bf k'}
a({\bf k})a^*({\bf k'}) 
\langle {\bf k'} |e^{i \hat{H}t} \hat{\bf k}  e^{-i \hat{H}t} |{\bf k} \rangle \right\} = \\
\\
\,\,\,\,\,\,\,\,\,\,\,\,\,\,\,\,\,\,\,\,\,\,\,\,\,\,\,\,\,\,\,\,\,\,\,\,\,\,\,\,\,\,\,\,\,\,\,\,\,\,\,\,\,\,\,\,\,\,\,\,\,\,\,\,\,\,\,\,\,\,\,\,\,\,\,\, \int d^D{\bf k} \int d^D{\bf k'}
a({\bf k})a^*({\bf k'}) 
\langle {\bf k'} \vert  \left[ \hat{\bf k},-ie{\bf E \cdot} \hat{{\bf r}}   \right]  \vert {\bf k} \rangle =- e{\bf E},  
\end{array}
\label{mom}
%\end{align}
\end{equation}
where $\hat{{\bf k}}$ and $\hat{{\bf r}}$ are respectively quantum mechanical momentum and coordinate operators.
From (\ref{mom}) it follows that  under the action of only the electric field the wave packet will accelerate indefinitely. This
 never happens in metals because of scatterings on  impurities, that randomly change the direction of motion. 
It is impossible then to trace trajectories of all wave packets and the natural language to 
describe such a system is provided by the semiclassical Boltzmann equation. 

In classical physics the Boltzmann equation is the evolution equation for the particle distribution in the phase space. We will always assume in this 
work that the system is spatially uniform on scales much larger than the distance between scatterers, where the classical Boltzmann equation
for scatterings on elastic impurities
 has the following form \cite{ashcroft-book,loss-03}
 
 \begin{equation}
  \frac{\partial f_{{\bf k}}}{\partial t} -e{\bf E} \frac{\partial f_{{\bf k}}} {\partial {\bf k}}
  = -\sum_{{\bf k'}}
   \omega_{{\bf k,k'}} (f_{{\bf k}}-f_{{\bf k'}}).
 \label{beint}
 \end{equation}

The rhs of (\ref{beint}) is called the collision term. For electrons that interact only with static impurities but non with each other the collision term
is a linear functional of the distribution function. This linearity
is not affected by the Pauli principle \cite{kohn-57}. However, when many body interactions contribute to the collision term the Pauli principle leads to
 contributions  proportional to 
$f_{{\bf k}}(1-f_{{\bf k'}})$ e.t.c. We will not consider the latter case in our discussion.
The scattering rate $ \omega_{{\bf k,k'}}$ depends on details of the scattering potential and should be found separately.
For a sufficiently smooth impurity, one can use wave packet equations to find the scattering cross-section by purely classical means but,
in most of realistic applications, a smooth potential approximation does not hold for an impurity.  Often the opposite limit of a
$\delta$-function type of a potential is considered as a reasonable assumption. This fact jeopardizes the
applicability of the semiclassical approach  but quantum mechanics provides a simple solution. There is the {\em  rule} that connects the quantum mechanical scattering matrix 
with the classical scattering rate. This rule is called the golden rule of quantum mechanics. For a weak impurity potential
   in the lowest Born approximation it reads \cite{perelomov-book}

 \begin{equation}
\omega_{{\bf k,k'}}=\frac{2\pi}{\hbar} |V_{{\bf k,k'}}|^2 \delta(\epsilon_{{\bf k}}-\epsilon_{{\bf k'}}),
\label{wt}
\end{equation}
where $\epsilon_{{\bf k}}$ is the kinetic energy of an electron with the momentum ${\bf k}$ and $V_{{\bf k,k'}}$ is the matrix element of the disorder potential between
two states of an electron before and after the scattering. In what follows we will assume that $\hbar=1$. The potential of  randomly placed impurities is $V({\bf r})=\sum_iv({\bf r}-{\bf R}_i)$, where $i$ enumerates impurities, ${\bf R}_i$ are their random positions and $v({\bf r}-{\bf R}_i)$
is the
potential of the single impurity with its center placed at ${\bf R}_i$. One can show \cite{kohn-57}, that for such a disorder $\langle |V_{{\bf k,k'}}|^2 \rangle_{dis} = n\vert v_{{\bf k-k'}}\vert^2$, where $n$ is the impurity concentration and
$v_{{\bf q}}$ is the Fourier transform of the single impurity potential at ${\bf R}_i=0$.

Together, the golden rule (\ref{wt}) and the classical Boltzmann equation (\ref{beint}) allow
to perform the quantitative
self-consistent calculation of the conductivity. Assume that the electric field is weak and look for the solution of the Boltzmann equation in the form

 \begin{equation}
 f_{{\bf k}}= f_{eq}(\epsilon_{{\bf k}})+ g_{{\bf k}},
\label{fg}
\end{equation}
where $f_{eq}(\epsilon_{{\bf k}})$ is the equilibrium distribution, which depends only on the energy of a particle and does not contribute to the current and
$g_{{\bf k}}$ is the correction linear in ${\bf E}$. At the steady state the term with the partial time derivative is zero and to linear order in ${\bf E}$ the correction
to the distribution satisfies the time-independent equation

  \begin{equation}
 -e{\bf E \cdot v_{{\bf k}}} \frac{\partial  f_{eq}(\epsilon_{{\bf k}})} {\partial \epsilon_{{\bf k}}}
  = -\sum_{{\bf k'}}
   \omega_{{\bf k,k'}} (g_{{\bf k}}-g_{{\bf k'}}),
 \label{beint1}
 \end{equation}
where in our case ${\bf v}_{{\bf k}}={\bf k}/m$.
Looking for the solution in the form $g_{{\bf k}}=g_0{\bf E\cdot k}$, one can find a self-consistent result
 \begin{equation}
g_{{\bf k}}=\frac{e\tau_{tr}{\bf E \cdot k}}{m} \cdot \frac{\partial  f_{eq}(\epsilon_{{\bf k}})} {\partial \epsilon_{{\bf k}}} ,
 \label{g1}
 \end{equation}
where 
 \begin{equation}
\frac{1}{\tau_{tr}}=\sum_{{\bf k'}}\omega_{{\bf k,k'}} (1- \cos ({\bf k,k'})),
 \label{g1}
 \end{equation}
 $\tau_{tr}$ is called the transport life time.
The electric current is given by the expression
 \begin{equation}
{\bf J}=-e\sum_{{\bf k}} g_{{\bf k}} {\bf v}_{{\bf k}}.
 \label{curr}
 \end{equation}
Using that at zero temperature $\frac{\partial  f_{eq}(\epsilon_{{\bf k}})} {\partial \epsilon_{{\bf k}}}=-\delta(\epsilon_F-\epsilon_{{\bf k}})$ one can arrive
at the following expression for the conductivity along the electric field in the 2D electron system
 \begin{equation}
\sigma_{xx}=\frac{e^2k_Fv_F\tau_{tr}}{4\pi m}.
 \label{currxx}
 \end{equation}

The important point is that the result (\ref{currxx}) is rigorous in the sense that when conductivity calculations are performed by formally exact quantum mechanical
techniques, such as the  summation of disorder averaged Feynman diagrams in the Kubo formula in the diffusive regime and disregarding 
higher order effects, such as
the weak localization, one arrives at the same quantitative result.

\subsection{Difficulties with semiclassical approach in application to the AHE.}
 
The above discussion of free electrons demonstrates that it is possible to derive transport coefficients using the classical Boltzmann equation. The power of this approach is in its transparency. It operates only with concepts that have simple classical interpretations.
There is no problem with the gauge invariance. 

Because of its simplicity this approach is ideal for introducing to the physics of the electron transport. Many concepts of solid state physics can be explained
with a sufficient rigor 
without using complicated Green function techniques. Also, the semiclassical approach  is needed to develop the scientific intuition about the model.
Having done calculations for a simple problem one can be interested in further more complicated phenomena. It is always
good to have a preliminary expectation about the final result. The semiclassical approach allows to make such an insight, while it is considerably 
harder with other techniques. 

  Certainly, the semiclassical approach has limitations but it should not be considered as valid only in the classical limit. It makes rigorous estimates even for
impurities that have no analog in classical physics, such as for  delta-function potentials. Hence, its domain of validity is  larger.
However, in applications to the AHE, the semiclassical theory faced with a number of complications. Sometimes it has been speculated that 
the AHE is a purely quantum mechanical phenomenon that cannot be explained by classical means \cite{kontani-07}. We will discuss later that it is not true but
first, I explain the main arguments that created this scepticism.

In real crystals electrons are not free particles. They interact strongly with the periodic potential of the lattice. It has been well known that 
the lattice periodicity does not result in random scatterings. The Bloch theorem guarantees that eigenstates of the electron Hamiltonian in a perfect
crystal  have the form

 \begin{equation}
\psi_{n{\bf k}}({\bf r},t)=\frac{1}{L^{D/2}}e^{i{\bf k \cdot r}-i\epsilon_{{\bf k}}t} u_{n{\bf k}}({\bf r}), 
 \label{blochwf}
 \end{equation}
where $u_{n{\bf k}}({\bf r}) $ is a periodic in the elementary unit cell function. 
Due to this simplification, one can describe other interactions, such as scatterings on lattice imperfections
with an effective Hamiltonian, not dealing directly with the lattice potential. The price for this  is that the wave function has generally
a nontrivial periodic part $u_{n{\bf k}}({\bf r}) $ and the dispersion is no longer quadratic, $\epsilon_{{\bf k}} \ne {\bf k}^2/(2m)$
(usually it is possible to approximate it near important symmetry points by a quadratic dependence  with a renormalized mass). The spectrum is also not everywhere continuous
and splits into bands. 

\begin{figure}
%\ifpdf
%\vskip -0.1in
%\fi
\includegraphics[width=8.5 cm]{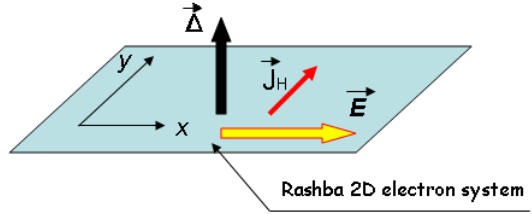}%{rashba24.eps} {Rashba24.pdf} 
\centering
\caption{\label{rashba} The 2D electron system described by the Hamiltonian (\ref{exch1}).}
\end{figure}

The multiple band structure  plays a very important role in the theory of the AHE. Because of it, operators of many observables
are matrices in the band index space, which can have nonzero off-diagonal (inter-band) elements. One of the simplest models of this kind, shown in 
Fig \ref{rashba}, is the Rashba coupled 2D electron system with an additional out of plane Zeeman interaction.
%%%%%%%%%%%%%%%%%%%%%%%%%%%%%%%%%%%%%%%%%%%%%%%%%%%%%%%%%%%%%%%%%%%%%%%%%%%%%%%%%%%%%%%
%%%%%%%%%%%%%%%%%%%%%%%%%%%%%%%%%%%%%%%%%%%%%%%%%%%%%%%%%%%%%%%%%%%%%%%%%%%%%%%%%%%%%%%%         
Its Hamiltonian reads
\begin{equation}
\hat{H}_{0}=\frac{k^{2}}{2m}+
\lambda (k_{y}\hat{\sigma}_{x}-k_{x}\hat{\sigma} _{y}) - \Delta \hat{\sigma}_{z},  
\label{exch1}
\end{equation}
where $\hat{\sigma}_i$ are Pauli operators. 
This Hamiltonian describes two bands with different dispersions 
\begin{equation}
\epsilon^{\pm}=\frac{k^2}{2m} \mp \sqrt{(\lambda k)^2+\Delta^2},
\label{disp1}
\end{equation}
where $+$/$-$ stand for major/minor band indexes.
The velocity operator $\hat{{\bf v}}=\partial \hat{H}_0/\partial {\bf k}$ has following components 

\begin{equation}
\hat{v}_x=\left(
\begin{array}{ll}
k_x/m   & i\lambda \\
-i\lambda   &  k_x/m
\end{array}
\right),\,\,\,\,\,\,\, 
\hat{v}_y=\left( 
\begin{array}{ll}
k_y/m   & \lambda \\
\lambda   &  k_y/m
\end{array}
\right).
\label{alphas}
\end{equation} 
It is straightforward to check that neither $\hat{v}_x$ nor $\hat{v}_y$ commute with the Hamiltonian.
This means that in the Bloch basis of eigenstates of the Hamiltonian (\ref{exch1}),
the velocity operator still has nonzero off-diagonal elements, for example,
\begin{equation}
\hat{v}_x=\left(
\begin{array}{ll}
v_x^{++}   & v_x^{+-} \\
v_x^{-+}   &  v_x^{--}
\end{array}
\right), \,\,\,\,\,\,\,\,\,\,v_x^{+-} \ne 0,  \,\,\,\,\,\,\,\,\,\,v_x^{-+} \ne 0 .
\label{vpm}
\end{equation}

The semiclassical interpretation of diagonal (intra-band) matrix elements of the velocity operator
is trivial. If one prepares a wave packet made of Bloch states of one band, the free motion of such a wave packet
will be the corresponding diagonal velocity. The off-diagonal velocity matrix elements are more subtle.
They do not affect the motion of a free wave packet. Only if a coherence among states of different bands is introduced
due to some perturbation can their expectation values become nonzero. This mixing can be produced e.g. by an
applied external electric field. Due to the Bloch vector dependence of the periodic part of the Bloch wave, the coordinate operator
generally has nonzero inter-band matrix elements
 \begin{equation}
\langle u_{n{\bf k}} |\hat{{\bf r}} |u_{\bar{n}{\bf k}} \rangle=\langle u_{n{\bf k}} \vert i\frac{\partial u_{\bar{n}{\bf k}}}{\partial {\bf k}} \rangle \ne 0,
 \label{bloch2}
 \end{equation}
where $n \ne \bar{n}$ and $\langle u_{n{\bf k}} \vert u_{n'{\bf k'}}\rangle $ is understood as the integral of the product of periodic parts of Bloch states over the unit cell or, in the case of the Rashba Hamiltonian 
(\ref{exch1}), it means the scalar product of spinor parts of  Hamiltonian eigenstates.

When the electric field is applied, the total Hamiltonian has a contribution $\hat{H}_E=e{\bf E \cdot \hat{r}}$, that not only accelerates wave packets,
 but also mixes states of different bands.
 Because of this, the expectation of inter-band parts of velocity
operator components becomes nonzero. In turn, this means that in the applied electric field the instantaneous velocity 
of the wave packet is no longer a corresponding diagonal
part of the velocity operator, but   contains an extra (anomalous) component. The situation reminds the chiral anomaly in the quantum field theory \cite{chiral-book} 
where the noncommutativity of the axial current operator with the Hamiltonian
leads to effects, unexpected from the Hamiltonian symmetries.

Similarly to the electric field, the impurity potential also mixes states of different bands. For example,  the point-like impurity potential $V({\bf r})=V_0 \delta({\bf r})$ 
in the chiral basis of the Rashba 2D electron gas has the following matrix form

\begin{equation}
\hat{V}_{{\bf k,k'}}=\frac{V_0}{L^2} \cdot\left(
\begin{array}{ll}
 \langle u^{+}_{{\bf k}} \vert  u^{+}_{{\bf k'}}  \rangle  &  \langle u^{+}_{{\bf k}} \vert  u^{-}_{{\bf k'}} \rangle \\
 \langle u^{-}_{{\bf k}} \vert  u^{+}_{{\bf k'}} \rangle  &   \langle u^{-}_{{\bf k}} \vert  u^{-}_{{\bf k'}}\rangle 
\end{array}
\right),
\label{delV}
\end{equation}
with nonzero off-diagonal matrix elements (see \cite{borunda-07} for explicit expressions).  
This means that the impurity role does not reduce to a simple instantaneous change of a direction of the particle  motion. When a wave packet
passes near such an impurity its wave function becomes distorted and the inter-band part of the velocity acquires  a nonzero expectation due to the local band mixing. Thus in the vicinity
of impurities wave packets move along unusual trajectories.

The AHE was found to be related exactly to that type of microscopic processes. Hence the construction of the semiclassical theory of this
effect faced with the problem of how to include the inter-band coherence into the purely classical description.

%%%%%%%%%%%%%%%%%%%%%%%%%%%%%%%%%%%%%%%%%%%%%%%%%%%%%%%%%%%%%%%%%%%%%%%%%%%%%%%%%%%%%%%%%%%%%%%%%%%%%%%%%%%%%%%%%%%%%%%%%%%%%%%%%%%%%%%%%

\section{Early theories of the AHE}
The first theoretical proposal to relate the AHE and the spin-orbit interaction 
was made by Karplus and Luttinger \cite{karplus-54}. They started from the fact that due to the relativistic corrections the 
effective Hamiltonian of an electron in a periodic lattice potential $V({\bf r})$ has an extra contribution due to the
spin orbit interaction
\begin{equation}
\hat{H}_{SO}=-\frac{1}{4m^2c^2} \hat{{\bf \sigma}} {\bf \cdot} ( {\bf p} \times {\bf \nabla}V).
\label{hso}
\end{equation}
This part of the Hamiltonian modifies  Bloch wave functions of electrons, introducing specific Bloch vector dependence in their 
periodic parts $u_{n{\bf k}}({\bf r})$. 
When the electric field is applied the corresponding term in the Hamiltonian 

\begin{equation}
\hat{H}_E=e{\bf E \cdot} \hat{{\bf r}}
\label{he}
\end{equation}
has nonzero matrix elements between states of different bands

\begin{equation}
\langle u_{n{\bf k}} \vert \hat{H}_E \vert u_{\bar{n}{\bf k}} \rangle=ie{\bf E \cdot} \langle u_{n{\bf k}} \vert
 \frac{\partial u_{\bar{n}{\bf k}}}{\partial {\bf k}} \rangle \ne 0,
\label{matH}
\end{equation}
where $\bar{n} \ne n$. This band mixing  ultimately leads to an unusual linear in the electric field
contribution to the transverse velocity.

 The spin orbit coupling alone does not lead to the AHE because the anomalous
transverse velocity, even if present, would have different signs in different degenerate bands unless the time reversal 
symmetry is broken. In ferromagnets this symmetry breaking appears spontaneously due to the exchange interaction,
 which is often approximated in theoretical models by the mean Zeeman-like field acting on electrons spins. In the simplest picture of the AHE
this Zeeman field creates a population imbalance between bands with opposite signs of the anomalous velocity thus leading to the
Hall current, proportional to the magnetization. This scenario is not always correct. For example, the recent research showed that often the strong AHE
appears near points of the spectrum where the magnetization field lifts degeneracies in the band structure \cite{nagaosa-rev}, leading to resonance features at
the Fermi energies in their vicinity.

Karplus and Luttinger did not expect that their theory would be the final answer to the AHE puzzle.
The reason was that the current they found was not gauge invariant, and thus could not describe the real 
observable. They pointed, however, that the magnitude of the found expression and its dependence on
the impurity concentration were in a good agreement with results of experimental measurements, thus suggesting that
their theory captured basic microscopic physics behind the AHE and the spin-orbit coupling and the magnetization must be the
                     forces, driving the effect.

Their theory was subsequently criticized by J. Smit \cite{smit-1,smit-2}, who made the first effort to design the 
gauge invariant theory of the effect. Smit agreed with that the spin orbit coupling is responsible for the AHE but
suggested a different mechanism. His approach was semiclassical in spirit. Smit's suggestion was to
look at the evolution of a wave packet, as a semiclassical object and to design a Boltzmann-like equation to describe the evolution
of the wave packet distribution function in the phase space. 
Today his name  is usually associated with only one of the contributions to the Hall current, called the skew scattering, which
was a new effect, not discussed by Luttinger at that time.
However, in addition to the idea of the skew scattering, Smit discussed other contributions. Tracing the evolution of a wave packet
he also found the anomalous velocity in the external field \cite{smit-2}, though
he did not think that this velocity contributes to the Hall conductivity because, he thought, it is exactly canceled by another effect. 

Smit pointed that the anomalous velocity follows from the change of the  polarization of wave packets by showing that
the average of the coordinate of the wave packet contains generally an additional component ${\bf A}$ due to the Bloch vector dependence of the periodic part
of the Bloch wave function
\begin{equation}
 {\bf r}_c ({\bf k},t)={\bf v}_{{\bf k}}t + {\bf A} = \frac{\partial \epsilon_{{\bf k}}}{\partial {\bf k}} t + \langle u_{{\bf k}}|
 i\frac{\partial }{\partial {\bf k}} u_{{\bf k}} \rangle.
 \label{chcen1}
 \end{equation}
 The second part ${\bf A}=  \langle u_{{\bf k}}|
 i\frac{\partial }{\partial {\bf k}} u_{{\bf k}} \rangle$ depends on ${\bf k}$. 
According to Smit, when the electric field accelerates the wave packet, the vector ${\bf k}$ changes and
consequently this changes the polarization ${\bf A}$. The wave packet becomes deformed.
This evolution of the polarization leads to an additional charge transport in the transverse to the electric field direction. Smit pointed that 
 ${\bf A}$ is not gauge invariant and rather its curl

\begin{equation}
{\bf F}=curl_{{\bf k}}{\bf A}
\label{curla}
\end{equation}
should enter the final result. In the modern terminology ${\bf A}$ and ${\bf F}$ are called the Berry connection and the Berry curvature respectively. 
%Although Smit did not name them this way, he was aware of their importance.

Smit's objection to the relevance of the anomalous velocity to the AHE conductivity also deserves a discussion. Smit pointed that in the DC limit wave packets cannot be constantly
accelerated. While the electric field changes the polarization by accelerating wave packets, scatterings on impurities produce on average the exactly 
opposite change of ${\bf k}$ if the system reaches the steady state. Thus Smit concluded that coordinate shifts at scatterings should have exactly
 opposite effect on  the wave packets polarization
and thus on the AHE conductivity. The coordinate shifts at scatterings, first introduced by Smit \cite{smit-1},
 indeed are the important part of the modern AHE theory. They were named 
"side-jumps" by Berger \cite{berger-1,berger-2,berger-3} who studied the effect in more details. 

Smit's work was the precursor of the modern semiclassical approach. He made the first effort to understand the anomalous Hall conductivity in classical
terms such as corrections to the velocity of wave packets, coordinate shifts at scatterings (side-jumps) and asymmetric scatterings at an impurity potential (skew scatterings).
All these ideas are currently incorporated in the theory, although his conclusion about the exact cancellation of the intrinsic and the side-jump contributions
is not supported by rigorous calculations. 

There are two main reasons why his arguments fail. One is that the "polarization" ${\bf A}$ is not a good quantum number and, in fact, is not
gauge invariant because it changes under an arbitrary momentum dependent change of the phase in the definition of Bloch states; 
%One can prepare a wave packet with the same average ${\bf k_c}$ but with nonzero angular momentum, which can be introduced by changing the 
%momentum dependent phase.
therefore one cannot apply classical balance arguments to it.
%Because of spin-orbit coupling such a wave packet will have a shifted average coordinate of the center of mass. 
%Scattered wave packet generally has nontrivial momentum dependent phase, that corresponds to changes in intrinsic angular momentum and thus even if momentum equilibrates in the DC limit
%The coordinate shifts due to scatterings and due to
%the acceleration in the electric field do not have to compensate each other because they differently 
%effect the orbital angular momentum. 
The second point, omitted by Smit, is that the side-jump can lead to the asymmetry of the
distribution function even without an asymmetry in the collision term kernel in the Boltzmann equation. Such a distribution asymmetry is rather due to the change of the kinetic energy that
particles experience after the side-jump in the presence of an electric field. The corresponding correction to the distribution function was named the anomalous distribution \cite{sinitsyn-07}.
When coupled to the conventional part of the velocity $\partial \epsilon_{{\bf k}}/\partial {\bf k}$ the anomalous distribution leads to the Hall current.

\subsection{Luttinger's rigorous theory: the quantum Boltzmann equation.}

In 1958 Luttinger published the detailed study of the AHE \cite{luttinger-58} based on the rigorous quantum mechanical approach
that he had designed with Kohn in a previous publication \cite{kohn-57}. Later this approach was generalized by Lyo and Holstein to
the regime of ac external fields \cite{holstein-ac}.  Luttinger's theory was correct but it did not find the general 
acceptance as a calculation tool and later many researchers have been looking for alternative techniques. The reason was 
that Luttinger's approach is very nontransparent. It involves many equations that self-consistently determine nongauge-invariant values.  

In this section I will try to make an explaining introduction to the Luttinger's paper. Rather than to follow directly his steps 
the goal here is to show  that it is possible to explain the Luttinger's derivation with a simple schema and a different notation, according to which it is easy
to classify AHE contributions in the Luttinger's approach and to make connections with the semiclassical theory. 

Luttinger starts with the evolution equation for the density matrix 
\begin{equation}
\frac{\partial\hat{\rho}}{\partial t}=i\left[ \hat{\rho},\hat{H} \right],
\label{qbe}
\end{equation}
where $\hat{H}$ is the Hamiltonian that includes both the disorder part and
the electric field

\begin{equation}
\hat{H}=\hat{H}_0+V(\hat{{\bf r}})+eE_x \hat{x}.
 \label{hh1}
\end{equation}

 In the stationary state one should require that 
\begin{equation}
\frac{\partial\hat{\rho}}{\partial t}=0.
\label{stst}
\end{equation}

If the solution of (\ref{qbe}) is found, the transverse electric current is given by the expression 

\begin{equation}
J_y=-e\Tr\left[ \hat{v}_y \hat{\rho} \right].
\label{ecurr}
\end{equation}
Hats mean that objects are matrices in the band index space.
Since the velocity operator $\hat{v}_y$ is diagonal in the momentum space, only the momentum-diagonal part of ${\hat{\rho}}$ is needed to 
calculate the current. However, $\hat{v}_y$ can have off-diagonal elements in the band index space.
From (\ref{qbe}), (\ref{stst}) Luttinger derives the analog of the Boltzmann equation, 
which contains terms that depend only on the diagonal in the momentum space  part of the density matrix. 
Schematically, it is useful to group terms that appear in his quantum Boltzmann equation as follows
\begin{equation}
E[DT](\hat{\rho}_{eq})-i[\hat{H}_0,\hat{\rho}_{neq}]=I_{col}(\hat{\rho}_{neq}),
 \label{qbe2}
\end{equation}
where 
$\hat{H}_0$ is the part of the Hamiltonian, independent of the electric field and of the disorder potential 
(Luttinger worked in the basis of Bloch states that diagonalize $H_0$).
$\hat{\rho}_{eq}$ is the equilibrium part of the density matrix (in the presence of the disorder but in the absence of the electric field) and
$\hat{\rho}_{neq}$ is the correction, linear in $E$. $[DT]$ means the "driving term" which explicitly couples to the electric field.
In the Bloch basis the driving term can be written as a series in powers of the disorder potential $V$, that
starts at $V^0$.

\begin{equation}
[DT](\hat{\rho}_{eq})=[DT]^{(0)}+V^2[DT]^{(2)}+\ldots.
\label{dt}
\end{equation}

The collision term for elastic scatterings on static impurities is linear in $\hat{\rho}_{neq}$ and 
also can be written as the series that starts at $V^2$.
\begin{equation}
I_{col}(\hat{\rho}_{neq})=V^2I_{col}^{(2)}(\hat{\rho}_{neq})+V^3I_{col}^{(3)}(\hat{\rho}_{neq})+V^4I_{col}^{(4)}(\hat{\rho}_{neq})+\ldots.
 \label{qbe3}
\end{equation}

This suggests to look for the solution for the nonequilibrium part of the density matrix in the form of a series in powers of $V$.
Separating terms of the same order in $V$ we find the chain of equations. The first equation allows to determine the largest term
in the expansion of $ \hat{\rho}_{neq}$ and others allow to express higher corrections through the lower ones.
From (\ref{qbe2}), (\ref{dt}) and (\ref{qbe3}) and the linearity of $I_{col}$ as a functional of $\hat{\rho}_{neq}$ it follows that
 this series begins at the term of order $V^{-2}$, i.e.
\begin{equation}
\hat{\rho}_{neq}= V^{-2}\hat{\rho}_{neq}^{(-2)}+V^{-1} \hat{\rho}_{neq}^{(-1)} + V^0 \hat{\rho}_{neq}^{(0)}+\ldots.
\label{exp3}
\end{equation}
%note that all terms in the series (\ref{exp3}) are of linear order in the electric field.

Simple power counting shows that to determine the correction of  order $V^{-2}$  it
 is enough to keep the driving term at zeroth order in $V$ and the first 
term in the collision part of  (\ref{qbe3}), i.e.

 \begin{equation}
E\left([DT]^{(0)}\right)_{diag}= \left( I_{col}^{(2)}( \rho_{neq}^{(-2)}) \right)_{diag},
 \label{qbe21}
\end{equation}
where the index $diag$ means that we take only the band-diagonal part of the expression. 
Luttinger found that $\hat{\rho}_{neq}^{(-2)}$ is diagonal in the band index and does not contribute to the Hall current. It, however, makes the dominating
contribution to the longitudinal current and is needed for further calculations.

Next order contribution $\hat{\rho}^{(-1)}$ satisfies the equation
\begin{equation}
I_{col}^{(2)}(\hat{\rho}_{neq}^{(-1})+I_{col}^{(3)}(\hat{\rho}_{neq}^{(-2)})=0,
\label{skew}
\end{equation}
where $\hat{\rho}^{(-2)}$ is already found by solving (\ref{qbe21}).

It turns out that $\hat{\rho}^{(-1)}_{neq}$, found from (\ref{skew}), is  still diagonal in band indexes and contains the antisymmetric contribution in the 
transverse to the
electric field direction. It leads to the transverse conductivity that, like $\hat{\rho}^{(-2)}_{neq}$, depends as $1/n$ on the impurity concentration $n$.

At zeroth (next) order in the disorder strength, both inter-band and intra-band
matrix elements become important.
One can separate 4 distinct parts.

\begin{equation}
\hat{\rho}^{(0)}_{neq} = \hat{\rho}_{int}+\hat{\rho}_{sj}+\hat{\rho}_{adist}+\hat{\rho}_{sk},
\label{rho0}
\end{equation}
where first two terms are purely off-diagonal and the other two are diagonal in band indexes. These contributions to the density matrix satisfy
following equations.
 \begin{equation}
E\left( [DT]^{(0)}\right)_{off-diag}-i[\hat{H}_0,\hat{\rho}_{int}]=0.
 \label{qbe22}
\end{equation}
Note that terms in (\ref{qbe22}) do not depend on the impurity potential. 

The equation for $\hat{\rho}_{sj}$ reads
\begin{equation}
-i[\hat{H}_0,\hat{\rho}_{sj}]=\left( I_{col}^{(2)}(\hat{\rho}_{neq}^{(-2)}) \right)_{off-diag}.
 \label{qbesj}
\end{equation}
$\hat{\rho}_{sj}$ is purely off-diagonal in band indexes and appears because the collision term 
$ I_{col}^{(2)}(\hat{\rho}_{neq}^{(-2)})$ has a nonzero off-diagonal part. 

The next contribution is diagonal and follows from the compensation between the higher order driving term and the collision part.
 \begin{equation}
E[DT]^{(2)}=I_{col}^{(2)}(\hat{\rho}_{adist}). 
 \label{qbe221}
\end{equation}

Finally, there is a contribution due to the compensation between two collision terms
 \begin{equation}
I_{col}^{(2)}(\hat{\rho}_{sk}) +I_{col}^{(4)}(\hat{\rho}_{neq}^{(-2)})=0.
 \label{qbesk}
\end{equation}

All four parts of the nonequilibrium density matrix in (\ref{rho0}) contribute to the Hall current via (\ref{ecurr}).
 Interestingly, since they are formally of zeroth power in $V$ and the  velocity
operator does not depend on $V$ the resulting conductivity due to these contributions for a Gaussian correlated disorder becomes independent of the 
impurity concentration.

\subsection{Noncommuting coordinates.}

Luttinger's theory is both complete and well justified quantum mechanically. However, 
his approach features the same problems as other rigorous quantum mechanical techniques in applications to the AHE.
It does not
explain what is  happening in simple terms. What can be concluded from his work is that
 off-diagonal elements of the density matrix and of the velocity operator play the important role.
Separately, they are not gauge invariant and only their 
product produced finally a gauge invariant current. Because of these complications, Luttinger did not discuss the semiclassical meaning of the derived 
contributions. 
% It is rather untransparent and it is hard 
%to say, based on its results what can possibly happen if extra interactions are added e.t.c.
  
To resolve this issue, in 1959 Adams and Blount made an effort to create the  semiclassical theory based on the introduction of
 noncommuting coordinates \cite{adams-59}. 

 The straightforward semiclassical approach, based on the preparation of a wave packet from states of the same band, may fail 
  when the electric field is applied. Since the electric field mixes states of different bands, a part of the initially free wave packet
  starts fast oscillations with frequencies $\omega \sim \epsilon_{n,{\bf k}}-\epsilon_{\bar{n},{\bf k}}$ in comparison to the rest of it.
  Such a wave packet does not satisfy the basic criteria of being a classical object because it would be composed of parts with 
strongly different oscillation frequencies.
 
  The resolution of this problem was first suggested by Adams and Blount \cite{adams-59}. 
  The off-diagonal part of the Hamiltonian due to the electric field can be considered as a periodic field that modifies the Bloch wave functions.
  Thus one can choose another Bloch basis, in which the term with the electric field has no inter-band matrix elements. To linear order in ${\bf E}=E_x\hat{{\bf x}}$ the periodic
part of such {\em modified} Bloch states reads

\begin{equation}
\vert u_{n{\bf k}}' \rangle=\vert u_{n{\bf k}}\rangle+ieE_x \sum \limits_{\bar{n}\ne n} 
\frac{\langle u_{\bar{n}{\bf k}}\vert\frac{\partial u_{n{\bf k}}}{\partial k_x}\rangle}{ \epsilon_{n{\bf k}}-\epsilon_{\bar{n}{\bf k}}} \vert u_{\bar{n}{\bf k}}\rangle,
 \label{bloch3}
 \end{equation}
 then  at $t=0$ one can prepare a new wave packet 

\begin{equation}
\Psi_{n{\bf k_c}}'({\bf r}) = 
 \int \frac{d^D{\bf k} }{L^{D/2}} a({\bf k})exp \left\{ i\left( {\bf k \cdot r}  \right)\right\} \vert u_{n{\bf k}}' \rangle,
\label{wwp12}
\end{equation}
which
in the external electric field does not split right away into differently oscillating components. It is now instructive to
calculate the velocity of the wave packet in the transverse to the electric field direction. Let the electric field points along x-direction and
$\hat{H}=\hat{H}_0+eE_x \hat{x}$ is the full Hamiltonian (we do not consider impurities yet). Then the wave packet evolves according to

\begin{equation}
\Psi_{n{\bf k_c}}'({\bf r},t) = e^{-i\hat{H} t}\Psi_{n{\bf k_c}}'({\bf r}).
\label{wwp11}
\end{equation}

The transverse velocity is calculated as follows
\begin{equation}
\begin{array}{l}
v_y=\frac{d}{dt} \langle \Psi_{n{\bf k_c}}'({\bf r},t) \vert \hat{y} \vert \Psi_{n{\bf k_c}}'({\bf r},t) \rangle =
\langle u_{n{\bf k}}' \vert -i\left[ i\frac{\partial}{\partial k_y},H_0 \right] \vert u_{n{\bf k}}' \rangle = \\
\\
=\frac{\partial \epsilon_{n{\bf k}}}{\partial k_y} +
ieE_x\left( \langle \frac{\partial u_{n{\bf k}}}{\partial k_y}\vert\frac{\partial u_{n{\bf k}}}{\partial k_x}\rangle-
\langle \frac{\partial u_{n{\bf k}}}{\partial k_x}\vert\frac{\partial u_{n{\bf k}}}{\partial k_y}\rangle
\right)=\frac{\partial \epsilon_{n{\bf k}}}{\partial k_y}-eE_xF_z^n,
\end{array}
\label{wwp2}
\end{equation}
where ${\bf F}^n$ is called the Berry curvature of the Bloch band with index $n$ and 
\begin{equation}
F_z^n=\Im \left( \langle \frac{\partial u_{n{\bf k}}}{\partial k_y}\vert\frac{\partial u_{n{\bf k}}}{\partial k_x}\rangle-
\langle \frac{\partial u_{n{\bf k}}}{\partial k_x}\vert\frac{\partial u_{n{\bf k}}}{\partial k_y}\rangle \right)
\label{fz}
\end{equation}
is its $z$-component.

The first term in the last line of (\ref{wwp2}) is just the usual velocity that equals the diagonal part of the velocity operator, while the rest of 
the expression is called the {\it anomalous velocity}. This anomalous contribution is ultimately responsible for the intrinsic AHE.

There is sometimes a misunderstanding about how the noncommuting coordinates appear in the theory. 
Sometimes it is simply stated that this happens to coordinate operators after the unitary transformation that switches to the basis of Bloch states. This, of course, is not true because
the unitary transformation alone cannot make commuting operators noncommuting. The reason is more subtle. 
In the standard Bloch basis with the periodic part $\vert u_{n{\bf k}} \rangle$ that diagonalizes $\hat{H}_0$ the coordinate operator has the form

\begin{equation}
\hat{{\bf r}} = i\frac{\partial}{\partial {\bf k}} + {\bf A}({\bf k}) + \hat{{\bf X}},
\label{coord}
\end{equation}
where ${\bf A}({\bf k})$ is diagonal in  band indexes and its elements are the Berry connections of Bloch bands and  $\hat{{\bf X}}$ is purely off-diagonal in band indexes.
%As we discussed, because the coordinate operator mixes different bands the full Hamiltonian $\hat{H}=\hat{H}_0+e{\bf E \cdot \hat{r}}$ also has off-diagonal matrix elements in that basis.
If we want to work with a modified Bloch band with periodic parts of Bloch states given in (\ref{bloch3}) we should switch to that new  modified
basis, i.e. to make the additional rotation
which transforms $\vert u_{n{\bf k}} \rangle$ into $\vert u_{n{\bf k}}' \rangle$. In that basis $\hat{H}_0$ is no longer pure diagonal and contains the off-diagonal component that  cancels with $e{\bf E}\cdot \hat{{\bf X}}$. Thus in the modified basis the full Hamiltonian has the form 
 
\begin{equation}
\hat{H} =\hat{H}_0'+e{\bf E \cdot} ( i\frac{\partial}{\partial {\bf k}} + {\bf A}({\bf k})),
\label{hamham}
\end{equation}
where now $\hat{H}_0'$ is, in fact, new operator, that 
has the same matrix form in the modified basis (\ref{bloch3}) as the old
free Hamiltonian $\hat{H}_0$ in the original Bloch basis $\vert u_{n{\bf k}}\rangle$.

In the basis (\ref{bloch3}) the Hamiltonian (\ref{hamham}) is by construction diagonal in band indexes including terms linear in the electric field.
Thus, when working with modified Bloch states, it is useful to regroup the terms  in the original Hamiltonian so that in the modified basis
it looks like we still have the unperturbed band-diagonal part $\hat{H}_0'$ but instead of the usual
coordinate operator the electric field couples only to its projection on the subspace of a given band.
The same observation holds if we consider the average of any operator over the state of the wave packet (\ref{wwp12}), prepared from the modified Bloch states
\begin{equation}
\langle \Psi_{n{\bf k_c}}({\bf r},t) \vert \hat{{\bf r}} \vert  \Psi_{n{\bf k_c}}({\bf r},t) \rangle \approx
i\delta'( {\bf k}-{\bf k_c}) + {\bf A}_n({\bf k_c}),
\label{rr}
\end{equation}
where we used the fact that the inter-band component of the coordinate operator has a zero expectation value and higher order terms in
the electric field can be disregarded.

Summarizing, if we consider the evolution of the wave packet made of modified Bloch states, mathematically, instead of
working with band mixing by the electric field, we  can assume that we still deal with the original free diagonal 
Hamiltonian $\hat{H}_0$ but  coordinate operators should be substituted by their projected versions

\begin{equation}
\hat{{\bf r}} \rightarrow \hat{{\bf r}}_c = i\frac{\partial}{\partial {\bf k}} + {\bf A}({\bf k}). 
\label{coord3}
\end{equation}

The anomalous velocity appears now as the result of the noncommutativity of such modified coordinate operators. Thus for the electric field along the x-axes
the transverse velocity contains a component
   \begin{equation}
   \begin{array}{l}
   \hat{v}_{y}^a=-i[\hat{y}_{c},eE_x\hat{x}_{c}]
   =-ieE_x \left[ i\frac{\partial}{\partial k_{y}} + A_{y}, i\frac{\partial}{\partial k_{x}} + A_{x} \right]=\\
\\
-eE_x \left( \frac{\partial A_{y}}{\partial k_{x}}
 -\frac{\partial A_{x}}{\partial k_{y}} \right) = -eE_xF_z.
   \end{array}
   \label{av2}
   \end{equation} 

Adams and Blount pointed that their approach can also be applied to the scattering problem on the impurity potential if the latter is sufficiently smooth
so that a moving in its field  wave packet feels only a weak gradient. One can truncate the 
off-diagonal part of the impurity potential $V({\hat{\bf r}}) \rightarrow V({\hat{\bf r}}_{{\bf c}}) $ so that the wave packet acquires an additional velocity

 \begin{equation}
 v^a_i= -i[\hat{r}_{c}^i, \hat{r}_c^j]\frac{\partial V}{\partial r_c^j}.
 \label{av4}
 \end{equation}
The comprehensive discussion of the noncommuting coordinates in Bloch bands can be found in \cite{blount-review}.
The theory of Adams and Blount was applied by Chazalviel \cite{chazalviel}, Nozieres,  Lewiner  \cite{nozieres}, Berger \cite{berger-1,berger-2,berger-3}
and by Lyo and Holstein \cite{holstein} to the AHE in III-V n-type semiconductors. Because of the degeneracy of electronic
bands in their model, they had to extend the theory
 to the case, which today is called "bands with a nonabelian Berry curvature". They started from the standard 8-band model with
a spin orbit interaction. Due to the hybridization the conducting bands are also
influenced by the spin orbit coupling. Similarly to the above discussion, Chazalviel and Nozieres projected coordinate operators to the space of only two conducting bands.
The problem, however, is that the latter have the same energy dispersions and thus the subsequent separation of conducting bands is impossible and one has to 
keep the off-diagonal elements of the coordinate operator in this subspace.
Thus the Berry connection and the Berry curvature become  $2\times 2$ matrices.

\begin{equation}
\hat{{\bf r}}_c=i\frac{\partial}{\partial {\bf k}} + \lambda \hat{{\bf s}}\times{\bf k},  
\label{rrrr}
\end{equation}
\begin{equation}
{\bf F}=2\lambda \hat{{\bf s}},  
\label{ffff}
\end{equation}
where $\hat{{\bf s}}$ is the pseudospin operator acting in the space of the conducting bands index.

This nonabelian case, however, is trivial because the Berry curvature does not depend on the Bloch vector. If one considers a 2D electron system, only the conserved out of plane component
of the pseudospin enters the Hamiltonian and the problem reduces to two separate bands with  the effective impurity potential
%%%%%%%%%%%%%%%%%%%%%%%%%%
\begin{equation}
V({\bf r}) \rightarrow V({\bf r})+\lambda \hat{{\bf s}} {\bf\cdot} \left({\bf k} \times {\bf \nabla} V({\bf r}) \right). 
\label{impso}
\end{equation}

The second term in (\ref{impso}) is functionally similar to the spin orbit coupling due to relativistic corrections (\ref{hso}) but with a different
strength $\lambda$, which can be considered as a renormalized spin orbit coupling constant. Formally,  corrections to the impurity potential and to the coordinate
operator (\ref{rrrr}) and (\ref{impso}) should be included in the full Hamiltonian of the 2D electron gas, together with the Rashba coupling. However, their effects are usually 
weak because of the weakness of the parameter $\lambda$ in (\ref{impso}), that makes the corresponding Hall conductivity also small \cite{chazalviel}. 
In contrast, the Hall effect due to the Rashba coupling
is nonperturbative. For example, the estimates of the intrinsic contribution for the Rashba coupled electron gas show the strength close to 
the universal value $e^2/2h$ in some range of parameters \cite{culcer-03}; so the effect of the Rashba term on the AHE is expected to dominate, 
although at large Fermi energies there 
can be a crossover between effects of two types of spin orbit couplings.

 The weakness of the Hall effect due to the coupling $\lambda$ in (\ref{impso}) 
can be understood in 
terms of topological defects in the band structure. Nonzero $\lambda$ is induced by 
a weak mixing of p-type orbitals to states of the conducting bands; hence it is suppressed,
 approximately inversely proportionally to the cubic power of the large gap
 between conducting and valence bands \cite{chazalviel}.  In contrast, the Rashba coupling,
even when small, creates the topological defect centered directly inside the conducting bands. At Fermi energies close to this "resonance" point the AHE is nonperturbative 
in the spin orbit coupling and hence is very strong.

Despite a partial success of the Adams's and Blount's approach, their semiclassical theory has strong limitations. One is the difficulty to apply it to a short range impurity potential. Such a potential 
destroys the wave packet and thus cannot be treated in a weak gradient approximation. The second problem is that this approach still operates with not strictly classical concepts such as noncommuting coordinates.
This complicates the interpretation of other objects in the semiclassical theory such as the distribution function. It also complicates
the derivation of the skew scattering contribution. Thus in all publications following this approach, the important 
part of the skew scattering contribution was missing. That part is parametrically similar to the side-jump contribution and cannot be disregarded. For example, the  Chazalviel's and  Nozieres's {\it et al} conclusion that the total impurity
concentration independent Hall conductivity
in their model
is the same in the magnitude but has opposite sign from the intrinsic contribution is wrong because of this omission.

\section{The Berry phase theory of the AHE.}

Sundaram and Niu \cite{sundaram-99} designed a very powerful and unifying framework to study the wave packet kinetics.
Their approach is based on the derivation of the effective Lagrangian of a wave packet moving in weak  fields.
The idea of the approach is that the time-dependent Schr\"odinger equation for a wave packet is realized from the variational principle with
the Lagrangian given by

\begin{equation}
L=\langle \Psi_{{\bf k_c}} \vert i\frac{d}{d t} - \hat{H} \vert \Psi_{{\bf k_c}} \rangle.
\label{lag1}
\end{equation}
The time dependence of the wave packet implicitly is contained in the time dependence of its average momentum ${\bf k_c}$ and coordinate 
${\bf r_c}=  \langle \Psi_{{\bf k_c}} \vert \hat{\bf r} \vert \Psi_{{\bf k_c}} \rangle$ and possibly in other explicitly time-dependent parameters in the system. 
 This allows to rewrite the time derivative in (\ref{lag1}) in terms of $\dot{\bf r}_{{\bf c}}$ and  $\dot{\bf k}_{{\bf c}}$

\begin{equation}
L={\bf k_c}\dot{\bf r}_{{\bf c}} - \epsilon({\bf k_c},{\bf r_c})+
\dot{\bf k}_{{\bf c}}\langle u \vert  \frac{\partial u}{\partial {\bf k_c}} \rangle + 
\dot{\bf r}_{{\bf c}}\langle u \vert  \frac{\partial u}{\partial {\bf r_c}} \rangle+
\langle u \vert  \frac{\partial u}{\partial t} \rangle,
\label{lag2}
\end{equation}
where $\vert   u \rangle$ is the unit cell periodic part of the Bloch wave.
%The energy $\epsilon({\bf k_c},{\bf r_c})$ is defined as
%\begin{equation}
%\fl
%\epsilon({\bf k_c},{\bf r_c})=\langle \Psi_{{\bf k_c}}  \vert \hat{H} \vert \Psi_{{\bf k_c}} \rangle \approx \langle \Psi_{{\bf k_c}} \vert  \hat{H}_0 \vert \Psi_{{\bf k_c}}\rangle+ \langle \Psi_{{\bf k_c%}} 
%\vert \Delta \hat{H} \vert \Psi_{{\bf k_c}} \rangle=\epsilon_c+\Delta \epsilon,
%\label{lag-en}
%\end{equation}
%where  
%\begin{equation}
%\langle \Psi_{{\bf k_c}}\vert  \hat{H}_0 \vert \Psi_{{\bf k_c}} \rangle= \epsilon_c({\bf k_c},{\bf r_c})
%\label{enc}
%\end{equation}
%is just the band energy at ${\bf k}={\bf k_c}$ and
%\begin{equation}
%\fl
%\begin{array}{l}
%\Delta \epsilon= \langle \Psi_{{\bf k_c}} \vert \Delta \hat{H} \vert \Psi_{{\bf k_c}}  \rangle= \\
%\\
%\langle \Psi_{{\bf k_c}} \vert 
%\frac{1}{2}\left( (\hat{{\bf r}}_{{\bf c}} -{\bf r_c}  )\cdot \frac{\partial \hat{H}_c}{\partial {\bf r_c}} + \frac{\partial \hat{H}_c}{\partial {\bf r_c}}\cdot
%(\hat{{\bf r}}_{{\bf c}} -{\bf r_c}  ) \right) \vert \Psi_{{\bf k_c}}  \rangle = -Im \langle \frac{\partial u}{\partial {\bf r_c}} \vert (\epsilon_c-\hat{H}_c) 
%\vert \frac{\partial u}{\partial {\bf k_c}} \rangle
%\end{array}
%\label{denc}
%\end{equation}
%is the correction due to potential gradients.

The first two terms in the Lagrangian (\ref{lag2}) are the same as in a typical Lagrangian of a classical particle with the classical mechanical Hamiltonian $H_{cl}=\epsilon({\bf k_c},{\bf r_c})$.  
The other terms are geometric, in the sense that their contribution to the action depends on the trajectory in the phase space but not on the rate of the motion along this trajectory. This is
the feature of the Berry phase in quantum mechanics and effects of the last three terms in (\ref{lag2}) can be called Berry phase effects.

The Lagrangian formulation provides a fully gauge invariant approach to the study of the wave packet dynamics. It is now easy to find the equations of motion in an 
arbitrary potential with a weak gradient
in a magnetic field ${\bf B}$ by varying the action over the trajectory. For the motion in a band with a dispersion $\epsilon_{{\bf k}}$ the result reads

\begin{equation}
\begin{array}{l}
\dot{\bf r}_{{\bf c}} =\frac{\partial \epsilon_{{\bf k_c}}}{\partial {\bf k_c}} - \dot{\bf k}_{{\bf c}} \times {\bf F},\\
\\
\dot{\bf k}_{{\bf c}}=-e{\bf E} -{\bf \nabla}V({\bf r_c}) -e \dot{\bf r}_{{\bf c}} \times {\bf B},
\end{array}
\label{wpeq}
\end{equation}
where ${\bf F}$ is the Berry curvature of the Bloch band. Wave packet equations show that the Berry curvature can be considered as an unusual magnetic field acting in the momentum space.
However, unlike the magnetic field in electrodynamics, which is a pure $curl$, the Berry curvature originates from a source. In electrodynamics, a magnetic field with such properties would
originate from magnetic monopoles. Their analogs in Bloch bands are the points of exact crossings of band dispersion curves \cite{blount-review, nagaosa-rev}.

 The wave packet equations can be generalized when the motion in degenerate bands is considered. If states of degenerate bands mix coherently the evolution becomes more complicated.
The state of the wave packet should then be considered having finite amplitudes in both bands \cite{culcer_a-06, ryuichi-05}

\begin{equation}
\vert \Psi_{{\bf k_c}} \rangle = \int d{\bf k} a({\bf k}) \left[ \eta_1 \vert \Psi_1\rangle +\eta_2 \vert \Psi_2\rangle \right],
\label{wpnab}
\end{equation}
where $\vert \Psi_i \rangle$ ($i=1,2$) are basis functions of the wave packets in each band. The coefficients $\eta_i$ enter the effective Lagrangian and 
should be considered as independent variables.
Their dynamics was found to be according to equations

\begin{equation}
i\frac{d\eta_i}{dt}=\left(H_{ij}-\dot{\bf k}_{{\bf c}}\langle u_i \vert i \frac{\partial u_j}{\partial {\bf k_c}}\rangle \right) \eta_j,
\label{wpnab}
\end{equation}
 where $H_{ij}$ are matrix elements of the Hamiltonian.

\subsection{The anomalous Nernst effect}
The recent triumph of the wave packet approach was the first semiclassical explanation and the quantitative theory of the intrinsic
anomalous Nernst effect (ANE), which is the AHE driven not by an external electric field but rather by the 
temperature gradient. The theory of the ANE \cite{xiao-06} is based on the previous understanding of  the intrinsic angular momentum \cite{sundaram-99,culcer-04} of a wave packet and 
also on
the observation \cite{xiao-05} that the wave packet equations (\ref{wpeq}) lead to the specific expression for the phase space volume. 
When considering the continuous limit it reads
\begin{equation}
\sum \limits_{{\bf k}} \rightarrow \int [d{\bf k}] \left( 1+e{\bf B\cdot F}  \right).
\label{vol}
\end{equation}
The physically measurable transport current is defined by 
\begin{equation}
{\bf j}={\bf J}-{\bf \nabla}\times {\bf M}({\bf r}),
\label{curM}
\end{equation}
where ${\bf J}$ is the microscopic current and  ${\bf M}({\bf r})$ is the magnetization density. The latter can be found from the grand canonical potential in the magnetic field
\begin{equation}
\fl
F=-\frac{1}{\beta} \sum \limits_{{\bf k}} \log \left( 1+e^{-\beta(\epsilon_M-\mu)} \right)=-\frac{1}{\beta}\int [d{\bf k}] \left( 1+e{\bf B\cdot F}  \right) \log \left( 1+e^{-\beta(\epsilon_M-\mu)} \right),
\label{FM}
\end{equation}
where $\beta=1/k_BT$ and $\epsilon_M=\epsilon({\bf k}) -{\bf m}({\bf k}){\bf \cdot B}$ is the electron energy in the magnetic field coupled to the magnetic moment
of a wave packet ${\bf m}({\bf k})=-i(e/2) \langle \nabla_{{\bf k}} u \vert \times [\hat{H}({\bf k}) -\epsilon_{{\bf k}}]  \vert \nabla_{{\bf k}} u \rangle $.

The magnetization is the magnetic field derivative of the thermodynamic potential
\begin{equation}
 {\bf M}({\bf r}) = -\left( \frac{\partial F}{\partial {\bf B}} \right)_{\mu,T}.
\label{mag}
\end{equation}
The expression for the intrinsic Hall current, induced by a weak temperature gradient 
originates from the phase space volume correction in (\ref{FM}).
Substituting (\ref{mag},\ref{FM})  into (\ref{curM}) one can find \cite{xiao-06}

\begin{equation}
\fl
{\bf j}_{int}=-e\frac{{\bf \nabla}T}{T} \times \int [d{\bf k}] {\bf F}  \left[ (\epsilon_M - \mu)f_0(\epsilon_{{\bf k}}) k_BT \log \left( 1+e^{-\beta(\epsilon_M-\mu)} \right) \right].
\label{FM1}
\end{equation}

\subsection{Wavepacket theory of intrinsic contribution and side-jump effect.}

The intrinsic contribution to the AHE is a straightforward consequence of the anomalous velocity term in wave packet equations (\ref{wpeq}). Under the action of the electric field,
 all the electrons in the Fermi sea will shift in the transverse direction, leading to the intrinsic Hall current \cite{jungwirth-02}  

\begin{equation}
{\bf J}^{int}=-e^2  {\bf E} \times \int [d{\bf k}] f_{\bf k} {\bf F}, 
\label{int5}
\end{equation}
where $f_{\bf k}$ is the electron distribution function in the given band and ${\bf F}$ is the  Berry curvature. 

Sinitsyn {\em et al} demonstrated \cite{sinitsyn-05} that wave packet equations can also be successfully applied to the problem of a scattering on an impurity,
 if the latter has a sufficiently smooth potential, thus providing the fully gauge invariant theory of the side-jump effect.
Integrating  (\ref{wpeq})
 over the time interval during which  a wave packet "feels" the impurity potential and assuming that this potential is sufficiently weak,
one can find a coordinate shift at a scattering event 
\begin{equation}
{\bf \delta  r}_{{\bf k',k}}^a=
{\bf F} \times ({\bf k'}-{\bf k}), 
\label{zse1}
\end{equation}
where ${\bf k}^{\prime }$ and ${\bf k}$ are center of mass momentums of the  wave packet respectively after and before the scattering.
This definition of the anomalous
coordinate shift at a scattering on an impurity (the side-jump) is different from some expressions suggested in the early theories. For example Berger \cite{berger-rev,engel-rev} assumed that 
$ \delta y^a  \sim  k_x$. The Berger's definition follows from the identification of the coordinate shift with the Berry connection.
Since the Berry connection is not gauge invariant, the old definition does not have a direct semiclassical meaning.
 In contrast, the expression (\ref{zse1}) as well as more general expressions (\ref{delr4}), (\ref{delr4b}) from  next section are gauge invariant and 
depend both on the in-going  ${\bf k}$ and the out-going ${\bf k'}$ Bloch vectors.
This difference becomes important when constructing the rigorous theory of the effect when the scattering on an impurity is not isotropic.

There are two main rather distinct effects due to  the anomalous shift. 
One is the side-jump accumulation. After the averaging over many scatterings, side jumps do not cancel and lead to the velocity renormalization by a correction
\begin{equation}
 v_{y}^{(sj)}({\bf k})=\sum_{\bf k^{\prime }}\omega_{{\bf k'}, {\bf k}}{\bf \delta  r}_{{\bf k',k}}^a. 
 \label{sum}
\end{equation}
The second effect is that when a scattering takes place in the presence of an external electric
field, there is a change in the potential energy upon a scattering given by 
\begin{equation}
\Delta U_{{\bf k',k}}=e{\bf E \cdot \delta r}^a_{{\bf k',k}}.
  \label{b1}
\end{equation}%
This change of energy  ultimately influences the Hall conductivity and should be properly included in the semiclassical Boltzmann equation.

According to \cite{sinitsyn-05}, the side-jump related conductivities depend only on parameters taken near the Fermi surface. This is in contrast to the intrinsic contribution that
depends on the integral of the Berry curvature over the whole Fermi sea. This may be one of the reasons why the comparison with experiments showed a good agreement with the intrinsic contribution
calculations that disregarded impurity effects, except keeping a finite life time of quasiparticles \cite{yao-04}. Although the intrinsic and the side-jump contributions to the transverse conductivity
do not depend on impurity concentrations and in this sense are parametrically similar,
still if the Berry curvature is weak near the Fermi surface but strong deep inside the Fermi sea, the intrinsic contribution can dominate.

 The ultimate example of this kind is the quantum anomalous Hall effect
that appears when the Fermi level is placed inside the gap in the bulk spectrum. 
In such insulators the gapless excitations are forbidden (except near the sample edges),
 so the side-jump and the skew scattering effects do not contribute to the conductivity  but if the band has a nonzero Berry curvature there is a quantized intrinsic contribution to the Hall current.
This happens e.g. in 2D Dirac bands, related to the graphene system \cite{sinitsyn-06,kane-05}.

%%%%%%%%%%%%

\section{Scattering rules.}

%We showed an example with wave packet that a proper redefinition of the velocity allows to treat the effect of band mixing with purely classical means.
%The only place where quantum mechanics was involved is the derivation of the expression for the Berry curvature, and the rest of the discussion does not envolve quantum mechanics.

%This will be 
The "philosophy" of the semiclassical approach is to operate only with classical concepts, however, using several rules that connect some of them with
purely quantum mechanical ones in order to achieve a quantitatively rigorous result.
 The expression for the Berry curvature is one of such rules, namely it relates the anomalous velocity to  Bloch wave functions. 
%Clearly, extra  rules are needed, for example because 
The scattering is described in quantum mechanics  by the scattering matrix, which has no analog in classical physics; therefore one should 
use the rules that connect the scattering matrix to the classical microscopic effects.

One such a scattering rule is widely known, and for its importance it is named the {\it golden rule} of quantum mechanics. 
Lets introduce a combined band-momentum index $l=(n,{\bf k})$.
The golden rule relates the
scattering rate  $\omega_{ll'}$ from the state $l'$ into the state $l$ in the continuous spectrum with the corresponding element of the scattering $T$-matrix 
\cite{perelomov-book}:
\begin{equation}
\omega_{ll'}=2\pi |T_{ll'}|^2 \delta(\epsilon_{l'}-\epsilon_{l}).
\label{WT}
\end{equation}
The $T$-matrix is defined as
\begin{equation}
T_{ll'}=\langle l| \hat{V}| \psi_{l'} \rangle,
\label{Tmatrix}
\end{equation}
where $\hat{V}$ is the impurity potential operator and 
$| \psi_{l} \rangle$ is the eigenstate of the full Hamiltonian 
$\hat{H}=\hat{H}_0+\hat{V}$ that satisfies the Lippman-Schwinger equation 
\begin{equation}
| \psi_{l} \rangle = |l\rangle  +\frac{\hat{V}}{\epsilon_{l}-\hat{H}_0+i\eta} | \psi_{l} \rangle. 
\label{psil}
\end{equation}

Scattering rates $\omega_{ll'}$ cannot include all the possible information encoded in the scattering matrix. This is obvious because entries of the $T$-matrix are complex numbers and entries 
of the matrix of scattering rates are real. In the golden rule only the absolute value of the $T$-matrix elements are represented. Thus the semiclassical approach, which uses only the golden
rule as relating the classical and the  quantum descriptions of the scattering, should generally fail.
 This is indeed the case in the AHE. The golden rule does not contain the information about the side-jump effect at a scattering
event. This fact forced authors of \cite{sinitsyn-sj}  to search for the gauge invariant expression for the side jump that would connect it to the scattering matrix.  

Such an expression indeed can be derived. In the lowest Born approximation it has a particularly simple form,
\begin{equation}
 \delta {\bf r}_{l'l} = \langle u_{l'}| i\frac{\partial}{\partial {\bf k'}} u_{l'} \rangle - 
 \langle u_{l}| i\frac{\partial}{\partial {\bf k}} u_{l} \rangle
 - \hat{{\bf D}}_{{\bf k',k}} arg(V_{ l',l} ),
\label{delr4}
\end{equation}
where ${\rm arg}[a]$ is the phase of the complex number $a$ and
$$
\hat{{\bf D}}_{{\bf k',k}}= \frac{\partial}{\partial {\bf k'}} + \frac{\partial}{\partial {\bf k}}.
$$
This type of expressions has been first found even before the work  \cite{sinitsyn-sj}, however, beyond the AHE theory. Belinicher {\em et al} derived it 
to apply in the photovoltaic 
effect \cite{belinicher-82}.
 They showed that when electrons absorb a polarized light they make shifts (\ref{delr4}), where $V$ would be responsible for the electron-photon interaction.  
Such shifts of the form (\ref{delr4}) finally contributed to the photo-induced conductivity in their model. Unfortunately the work \cite{belinicher-82} had been unnoticed by the modern Hall effect community
until the
expression (\ref{delr4}) was independently rederived in \cite{sinitsyn-sj} and applied to the AHE-problem.

The expression for the side jump (\ref{delr4}) is gauge invariant. Interestingly, it restores the information, lost in the golden rule. Unlike the golden rule that in the lowest Born 
approximation depends on the absolute value of the scattering potential, the coordinate shift expression depends on its phase but does not depend on its absolute value.
 Thus the 
expression (\ref{delr4}) can be considered as  complimentary to the golden rule.

 If the impurity potential is spin independent, then the side-jump
does not depend explicitly on the type of the impurity potential and can be expressed in terms of  
initial and final states only \cite{sinitsyn-sj,sinitsyn-07}:
\begin{equation}
 \delta {\bf r}_{l'l} = \langle u_{l'}| i\frac{\partial}{\partial {\bf k'}} u_{l'} \rangle - 
 \langle u_{l}| i\frac{\partial}{\partial {\bf k}} u_{l} \rangle
 - \hat{{\bf D}}_{{\bf k',k}} {\rm arg}[\langle u_{l'}|u_{l}\rangle].
\label{delr4b}
\end{equation}
At a weak scattering angle it reduces to equation (\ref{zse1}), derived in previous section \cite{sinitsyn-06}.
The derivation of the golden rule (\ref{WT}) and the expression for the coordinate shift (\ref{delr4b}) can be done by considering a scattering of a wave packet.
Imagine a state, that initially coincides with the Bloch state $\psi_{l}({\bf r},t)$
 under the  influence of a weak potential of an impurity $V({\bf r})$.
The solution of the time-dependent Schr\"odinger equation can be written in terms of the eigenvectors $\psi_{l'}({\bf r},t)$ of the unperturbed Hamiltonian  
\begin{equation}
\psi^{out}_{l}({\bf r},t)=\sum_{l'} C_{l'}(t) \psi_{l'}({\bf r},t).
\label{split}
\end{equation}
Consider the wave packet, that was initially gathered around the state $l$. Then
in the lowest order in the strength of the potential the perturbation theory leads to the following expression for time-dependent coefficients 
$ C_{l'}(t)$ (see Eq. 19.9 in \cite{scatt-book})
\begin{equation}
 C_{l'}(t)=-iV_{l'l} \int_{-\infty}^t e^{i(\epsilon_{l'} -\epsilon_l)t'} dt' + \delta_{l'l} .
\label{ckt}
\end{equation}
The higher order terms can be incorporated into the above formula by merely substituting the 
$T$-matrix instead of the disorder matrix elements (see Eq. 19.10 in \cite{scatt-book}).
\begin{equation}
 C_{l'}(t)=-iT_{l'l} \int_{-\infty}^t  e^{i(\epsilon_{l'} -\epsilon_l)t'} dt' + \delta_{l'l}. 
\label{ckt1}
\end{equation}
%The integral in (\ref{ckt}) is formally diverging, reflecting the fact that for infinite interaction time the initial state is completely destroyed.

From this solution one can show that  for $l' \ne l$,  $ C_{l'}(t)$ change with time according to the law 

\begin{equation}
\frac{d| C_{l'}(t)|^2}{dt}=2\pi\vert T_{l'l} \vert^2 \delta (\epsilon_{l'} -\epsilon_l).
\label{chch}
\end{equation}
The coefficient $| C_{l'}(t)|^2$ has the meaning of the probability of the electron to be in the state $l'$, from which immediately  the golden rule (\ref{WT}) follows.

The coordinate shift expression was derived in a similar fashion. One can prepare a wave packet, approaching the impurity at the point ${\bf r}_{imp}=0$ according to the law 
${\bf r}(t)={\bf v}_{0l}t $ at $t \rightarrow -\infty$. The scattering on an impurity generally destroys the wave packet, however having the scattering matrix one can 
write a formal expression for the average coordinate of the state of the wave packet after the scattering. According to \cite{sinitsyn-sj} it can be written in the form

\begin{equation}
\fl
{\bf r_{+\infty}}=\int d{\bf r} \left( \Psi^{out}({\bf r},+\infty)\right)^* {\bf r} \Psi^{out}({\bf r},+\infty) = 
\sum_{l'} |C_{l'}(t\rightarrow +\infty)|^2 \left(   {\bf v}_{0l'}t + \delta{\bf r}_{l'l}  \right).
\label{finr}
\end{equation}
Since $|C_{l'}(t\rightarrow +\infty)|^2$ has the classical meaning of the scattering probability into the state $l'$, the expression in the parentheses is reasonable to interpret as the 
coordinate of the particle if it is scattered in the state $l'$. Then the term $ {\bf v}_{0l'}t$ simply tells that after the scattering 
the particle moves with the new velocity ${\bf v}_{0l'}$ and thus the expression $\delta{\bf r}_{l'l}$ in (\ref{finr}) can be interpreted as the coordinate shift at a scattering event.
This identification leads finally to the gauge invariant expression (\ref{delr4}).

%For weak disorder one can approximate the scattering state $| \psi_{l} \rangle$ 
%by a truncated series in powers of $V_{ll'}=\langle l|\hat{V} |l'\rangle$ and
%take $l=({\bf k},\mu)$ as the combined (momentum,band) index of the eigenstate $|l\rangle=|{\bf k},\mu\rangle$ of
%the unperturbed  Hamiltonian $\hat{H}_0$: 
%\begin{equation}
%| \psi_{l} \rangle \approx |l\rangle +\sum_{l''} \frac{V_{l''l}} {\epsilon_{l}-\epsilon_{l''}+i\eta} | l''\rangle + \ldots
%\label{ser1}
%\end{equation}
%For example the $T$-matrix up to the second order in $\hat{V}$ is

The side-jump is a weak effect and thus its expression in the lowest Born  approximation is sufficient for the further discussion. However the lowest Born approximation in the golden rule
is inappropriate for the theory of the AHE.
For a weak disorder one can use the expression of the $T$-matrix in terms of the Born series in powers of disorder potential matrix elements
\begin{equation}
T_{ll'}\approx V_{ll'}+\sum_{l''} \frac{V_{ll''}V_{l''l'}}{\epsilon_{l'}-\epsilon_{l''}+i\eta}+ \ldots.
\label{Tmatrix2}
\end{equation}
One can consider only several first terms in this series in order to capture the basic microscopic processes.
Substituting Eq. (\ref{Tmatrix2}) into Eq. (\ref{WT}) one will arrive at the expansion 
 \begin{equation}
 \omega_{ll'}=\omega^{(2)}_{ll'}+\omega^{(3)}_ {ll'}+\omega^{(4)}_ {ll'}+\cdots,
 \label{om1}
 \end{equation}
where
\begin{equation}
\omega^{(2)}_{ll'}=2\pi \langle |V_{ll'}|^2 \rangle_{dis} \delta (\epsilon_{l} -\epsilon_{l'}),
\label{om2}
\end{equation}
\begin{equation}
\omega^{(3)}_ {ll'}=2\pi \left ( \sum_{l''} \frac{\langle V_{ll'}
 V_{l'l''} V_{l''l}\rangle_{dis}}
{\epsilon_{l} -\epsilon_{l''}-i \eta} +c.c. \right) \delta (\epsilon_{l} -\epsilon_{l'}),
\label{om3}
\end{equation}
and so on. 
The skew scattering contribution to the Hall effect follows from the antisymmetric part of the scattering rate
\begin{equation}
\omega_{ll'}^{(a)} \equiv \frac{\omega_{ll'} - \omega_{l'l}}{2}.
\label{ss1}
\end{equation}
%NIK: Changed l'l to ll'
Since $\omega^{(2)}_{ll'}$ is symmetric, the leading contribution to $\omega_{ll'}^{(a)}$ appears at order $V^{3}$, at which
 the antisymmetric part of the scattering rate is particularly simple \cite{luttinger-58,leroux-72}
\begin{equation}
\begin{array}{l}
\omega^{(3a)}_ {ll'}= -(2\pi)^2  \sum \limits_{l''} \delta (\epsilon_{l}
 -\epsilon_{l''}) {\rm Im} \langle V_{ll'} V_{l'l''} V_{l''l}\rangle_{dis} 
 \delta (\epsilon_{l} -\epsilon_{l'}),
\end{array}
\label{omasym1}
\end{equation}
with the superscript $3a$ meaning that this is the antisymmetric part of the scattering rate calculated 
at order $V^3$.
Usually,  properties of the skew scattering were inferred only from this lowest order antisymmetric part of $\omega_{ll'}$. Thus it is customarily assumed that
$\omega^{(a)}_ {ll'}$ is proportional to the impurity concentration. This, however, holds only in the lowest nonzero order, i.e. for $\omega^{(3a)}_ {ll'}$. 
In the next order the antisymmetric scattering is proportional 
to the product of four disorder vertexes. For a Gaussian correlated potential 
$\langle V\cdot V\cdot V\cdot V \rangle_{dis} \sim \langle V\cdot V \rangle_{dis} \langle  V\cdot V \rangle_{dis}  \sim n^2$, where $n$ is the impurity concentration. Thus the higher order contribution behaves as
$\omega^{(4a)}_ {ll'}\sim n^2$, unlike $\omega^{(3a)}_ {ll'} \sim n$. This means that the higher order term is different parametrically and has a rather distinct microscopic origin; therefore
it should not be disregarded. 
Moreover, the contribution to the conductivity, arising from $\omega^{(4a)}_ {ll'}$, is parametrically similar to the side-jump related contributions. Hence, to derive all important effects one should calculate the symmetric part of $\omega_ {ll'}$ in the lowest Born approximation and then calculate its antisymmetric part, including next {\it two} orders in $V$.

 \begin{equation}
 \omega_{ll'} \approx \omega^{(2)}_{ll'}+\omega^{(3a)}_ {ll'}+\omega^{(4a)}_ {ll'}.
 \label{om111}
 \end{equation}

\section{Mechanisms  of  the AHE in the semiclassical Boltzmann equation.}
Eqs. (\ref{WT}) and (\ref{delr4b}) contain the quantum mechanical information necessary to achieve  quantitatively rigorous predictions working in the 
framework of the semiclassical Boltzmann
equation. This approach takes into account both the change of the direction and the coordinate shift during a scattering in a homogeneous crystal 
in the presence of a driving electric field $\bf E$. Keeping only terms up to the linear order in the electric field the Boltzmann
equation reads \cite{sinitsyn-07}
\begin{equation}
  \frac{\partial f_l}{\partial t} -e{\bf E}\cdot{\bf v}_{0l} \frac{\partial f_{0} (\epsilon_l) }{\partial {\epsilon_l}}
  = -\sum_{l'}
   \omega_{ll'} \left[ f_{l}-f_{l'}-\frac{\partial f_{0} (\epsilon_l) }{\partial {\epsilon_l}}e{\bf E} \cdot \delta {\bf r}_{ll'} \right], 
\label{beintt}
 \end{equation}
 where expressions for $\omega_{ll'}$ and $\delta {\bf r}_{ll'}$ were derived in previous section,  ${\bf v}_{0l}$ is the usual velocity 
\begin{equation}
{\bf v}_{0l} = \partial \epsilon_l/\partial {\bf k}
\label{vol12}
\end{equation}
and  $f_{0}(\epsilon_l)$ is  the equilibrium distribution.

%NIK: I split paragraph here into two.
The Boltzmann-type equation (\ref{beintt}) has the standard form, as (\ref{beint}) for free electrons, except 
 the allowed inter-band transitions
and the coordinate shift effect which 
is taken into account in the last term in the collision integral on the rhs of Eq. (\ref{beintt}). The derivation of this term has been explained in 
\cite{sinitsyn-05,sinitsyn-06,sinitsyn-07}. It follows from the fact that under the scattering in the electric field from the state $l'$ into the state $l$, the side-jump is associated with the change of the
potential energy $\Delta U_{ll'}=e{\bf E} \cdot \delta {\bf r}_{ll'}$, which has to be compensated by the change of the kinetic energy. Thus the conservation of energy requires that $\epsilon_{l'}-\epsilon_l=e{\bf E} \cdot \delta {\bf r}_{ll'}$.
This in turn means that the equilibrium distribution does not annihilate the collision term anymore because $f_0(\epsilon_l)-f_0(\epsilon_{l'}) \approx -\frac{\partial f_{0} (\epsilon_l) }{\partial {\epsilon_l}}e{\bf E} \cdot \delta {\bf r}_{ll'}$, which
is exactly the last term in (\ref{beintt}).  The validity of these arguments was confirmed in numerical simulations \cite{sinitsyn-05}.

The total distribution function $f_{l}$ in the steady state ($\partial f_l/\partial t =0$) 
can be written as
\begin{equation}
f_{l}=f_{0}(\epsilon_l)+g_{l}^s+g_l^{a1}+g_l^{a2}+g_l^{adist},
\label{ffgg}
\end{equation}
where   $g_l^s$, $g_l^{a1}$, $g_l^{a2}$ and $g^{adist}$ are non-equilibrium corrections to the distribution function of linear order in the electric field
(the label {\em adist} stands for the {\it anomalous distribution}).
They solve self-consistent time-independent equations:\cite{sinitsyn-07}
\begin{equation}
- e{\bf E}\cdot{\bf v}_{0l} \frac{\partial f_{0} (\epsilon_l) }{\partial {\epsilon_l}}
  = -\sum_{l'} \omega_{ll'}^{(2)} (g_{l}^s-g_{l'}^s ),
\label{bbeint3}
\end{equation}
\begin{equation}
\sum_{l'} \omega_{ll'}^{(3a)} (g_{l}^{s}-g_{l'}^s )+\sum_{l'} \omega_{ll'}^{(2)} (g_{l}^{a1}-g_{l'}^{a1} )=0,
\label{bbeint3a}
\end{equation}
\begin{equation}
\sum_{l'} \omega_{ll'}^{(4a)} (g_{l}^{s}-g_{l'}^s )+\sum_{l'} \omega_{ll'}^{(2)} (g_{l}^{a2}-g_{l'}^{a2} )=0
\label{bbeint4a}
\end{equation}
and
\begin{equation}
\sum_{l'}\omega_{ll'}  \left( g^{adist}_l - g^{adist}_{l'} -\frac{\partial f_0(\epsilon_l)}{\partial \epsilon_l} e{\bf E} \cdot  \delta {\bf r}_{ll'} \right) =0.
\label{ganl}
\end{equation}
One can deduce the dependence of the distribution corrections on the impurity concentration by noticing that $\omega_{ll'}^{(2)} \sim n$, then
from (\ref{bbeint3}) follows that $g_l^s \sim n^{-1}$. Then from (\ref{bbeint3a}) and the fact that $\omega_{ll'}^{(3a)} \sim n$
 follows that $g_l^{3a} \sim n^{-1}$, then from $ \omega_{ll'}^{(4a)} \sim n^2$ and (\ref{bbeint4a}) follows that $g_l^{4a} \sim n^0$
and from (\ref{ganl}) follows $g^{adist}_l\sim n^0$. A detailed solution of equations 
(\ref{bbeint3}-\ref{ganl}) in a 
special model was demonstrated in \cite{sinitsyn-07}.

The first correction $g_l^s$ is symmetric and others $g_l^{adist}$, $g_l^{a1}$ and $g_l^{a2}$ are asymmetric in the transverse to the electric field direction.
When coupled to the usual part of the velocity the latter lead to 3 separately gauge invariant contributions to the conductivity. For the electric field along the 
$x$-axes these read

\begin{equation}
\sigma_{yx}^{adist} =- e \sum_l( g_l^{adist}/E_x) (v_{0l})_y \sim n^0,
\label{jtotal_adist}
\end{equation}
\begin{equation}
\sigma_{yx}^{sk1} = -e \sum_l( g_l^{a1}/E_x) (v_{0l})_y \sim n^{-1},
\label{jt1}
\end{equation}
\begin{equation}
\sigma_{yx}^{sk2} = -e \sum_l( g_l^{a2}/E_x) (v_{0l})_y \sim n^{0}.
\label{jt2}
\end{equation}
The last two contributions can be called the skew scattering conductivities because they originate from the asymmetric part of the collision term kernel $\omega_{ll'}$. 
It is, however, reasonable to distinguish between the two because of their different dependence on the impurity concentration. The first
skew scattering conductivity (\ref{jt1}) is  the {\em conventional skew scattering}, that has been discussed by many authors \cite{luttinger-58,leroux-72}. The second one (\ref{jt2})
was generally discarded, although it is parametrically the same as the side-jump conductivity. The explicit quantitative estimates of 
$\sigma_{yx}^{sk2}$ so far exist only for the massive 2D Dirac band \cite{sinitsyn-07}. Because of the lack of the proper terminology we will
call (\ref{jt2}) the {\em intrinsic skew scattering} because this conductivity, similarly to the intrinsic contribution, is independent of the 
impurity concentration $n$.

The asymmetric corrections to the distribution do not exhaust all mechanisms of the AHE.
We already discussed that between scatterings under the action of the electric field, wave packets move with an extra velocity 
\begin{equation}
{\bf v}_l^a=  e {\bf E} \times{\bf F}_l.
\label{tovel}
\end{equation}
This velocity is linear in the electric field, therefore we did not consider its effect on the distribution function. However, 
when coupled to the equilibrium part of the distribution it produces a finite Hall current.
\begin{equation}
\sigma_{yx}^{int} = e^2 \sum_l f_0(\epsilon_l) F_l \sim n^0, 
\label{jtotal_i}
\end{equation}

Finally, the  accumulation of coordinate shifts after many scatterings can be
described semiclassically on average as an additional velocity contribution

\begin{equation}
{\bf v}_l^{sj}=  \sum_{l'} \omega_{l'l} \delta {\bf r}_{l'l},
\label{tovel}
\end{equation}
\begin{equation}
\sigma_{yx}^{sj} = - e\sum_l (g_l^s/E_x) \left( \sum_{l'} \omega_{l'l} (\delta {\bf r}_{l'l})_y  \right) \sim n^0.
\label{jtotal_sj}
\end{equation}

Thus the total transverse conductivity can be written as the sum of five contributions:
\begin{equation}
\sigma_{yx}^{total}=\sigma_{yx}^{int}+\sigma_{yx}^{adist}+\sigma_{yx}^{sj}+\sigma_{yx}^{sk1}+\sigma_{yx}^{sk2}.
\label{jttt}
\end{equation}

\subsection{Semiclassical versus fully quantum mechanical techniques.}
Besides the semiclassical theories,
the research on diluted magnetic semiconductors also stimulated the theoretical interest in other quantitative approaches. Due to the relative simplicity of several 
important models, such as the Rashba coupled 2D electron system, a number of publications appeared recently with rigorous quantum mechanical calculations by Kubo and Kubo-Streda formulas 
\cite{dugaev-05, inoue-06, nunner-07, sinitsyn-06, sinitsyn-07, dugaev-01,bruno-dirac,kontani-94,kontani-97} and by a variety of the quantum Boltzmann equation
and the Keldysh techniques \cite{liu-05, liu-06,onoda_2-06,culcer-06}.
Sinitsyn \etal \cite{sinitsyn-sj,sinitsyn-07}  demonstrated
 the 1-1 correspondence between semiclassical contributions to the AHE
and the summation of relevant subseries of Feynman diagrams in the Kubo-Streda formula \cite{streda}. Similar agreement was established with
the Luttinger's theory \cite{luttinger-58}. 
The results are summarized in Table \ref{AHE-table}.  

According to \cite{sinitsyn-07}, the classification of contributions in the Kubo formula is not merely by a separation of
diagrams into the disorder free part and the vertex correction but rather by the parts of the velocity matrices in chiral basis that stay inside the trace of the Kubo formula. Thus the intrinsic contribution
appears from the summation of all diagrams with only off-diagonal parts of the velocity vertexes in Bloch basis.  From this point of view the skew scattering is the most "classical" because it is due to the summation of all diagrams with only diagonal parts of velocity operators and the difference between
conventional and intrinsic skew scatterings is due to different types of disorder vertexes involved. The conventional skew scattering 
is due to the vertex correction that involved correlators of three or 
more disorder vertexes while the intrinsic skew scattering is due to only Gaussian disorder correlations. The side jump and the anomalous distribution effects are related to Feynman diagrams that contain one off-diagonal (inter-band)
and one diagonal (intra-band) parts of the velocity operator. This reflects the fact that although the side-jump itself is related to the anomalous velocity it contributes to 
the final current only after the electric field distorts the 
distribution function by a simple acceleration, i.e. by the coupling to the usual velocity in the Boltzmann equation or coordinate shifts create
 an anomalous distribution that again contributes to the final current via the coupling to the usual velocity.

\begin{table}
\fl
\caption{\label{AHE-table}Mechanisms of the AHE in the semiclassical Boltzmann equation (SBE),  the Luttinger's quantum Boltzmann equation (QBE) and the Kubo-Streda 
\cite{streda} formula (KSF). $\hat{v}^d_{x/y}$ and $\hat{v}^{od}_{x/y}$ stand for diagonal and
off-diagonal parts of the velocity operator in the Bloch state basis, $n$ is the impurity concentration, Tr is the summation over all states in the momentum space and over
all band indexes, $G^R,G^A$ are respectively  retarded and advanced Green functions of the electron system and other symbols have the same meaning as 
in the bulk of the text.}
\begin{indented}
\lineup
\item[]\begin{tabular}{@{}*{7}{l}}
\br                              
mechanism &strength&SBE&QBE&KSF \cr 
of the AHE&of $\sigma_{yx}$& $E_x\sigma_{yx}$ &  $E_x\sigma_{yx}$ &      $\sigma_{yx}$       \cr 
\mr
intrinsic&$O(n^0)$&$-e\Tr(f_0v^a_y)$  &$-e\Tr(\hat{\rho}_{int}\hat{v}^{od}_y)$&$\sigma^{II}_{yx}+\sigma^{I,int}_{yx}$\cr\\
side-jump&$ O(n^0)$&$-e\Tr(g^sv^{sj}_y)$&$-e\Tr(\hat{\rho}_{sj}\hat{v}^{od}_y)$&$\frac{e^2}{2\pi}\Tr\left(\hat{v}_y^{od}\hat{G}^R\hat{v}_x^d \hat{G}^A\right)$\cr
accumulation&      &    &                                 &\cr \\
anomalous &$ O(n^0)$&$-e\Tr(g^{adist}v_{0y})$ &$-e\Tr(\hat{\rho}_{adist}\hat{v}^{d}_y)$&$\frac{e^2}{2\pi}\Tr\left(\hat{v}_y^{d}\hat{G}^R\hat{v}_x^{od} \hat{G}^A\right)$\cr 
distribution\\
\\
conventional    &$ O(n^{-1})$&$-e\Tr(g^{a1}v_{0y})$  &$-e\Tr(\hat{\rho}^{(-1)}\hat{v}^{d}_y)$&$\frac{e^2}{2\pi}\Tr\left(\hat{v}_y^{d}\hat{G}^R\hat{v}_x^d \hat{G}^A\right)$\cr 
 skew scattering&       &($\omega_{{\bf k,k'}}^a\sim n$)  &                       &(nongaussian vertex)\cr \\
\\
intrinsic&$ O(n^0)$& $-e\Tr(g^{a2}v_{0y})$ &$-e\Tr(\hat{\rho}_{sk}\hat{v}^{d}_y)$&$\frac{e^2}{2\pi}\Tr\left(\hat{v}^{d}_y\hat{G}^R\hat{v}_x^d \hat{G}^A\right)$\cr
skew scattering &       &( $\omega_{{\bf k,k'}}^a\sim n^2$ ) &                       &(gaussian vertex)\cr \\
\br
\end{tabular}
\end{indented}
\end{table}

\subsection{Terminology in the AHE theory.}
After the above discussion of all the effects leading to the AHE it seems appropriate to reconcile some of the differences 
in the terminology in 
recent publications. Generally it is stated that there are three basic microscopic effects, leading to the Hall current: the intrinsic contribution, 
the side-jump and the skew scattering. Table \ref{AHE-table} shows that this classification is
too restrictive. There are, in fact, five separately gauge invariant contributions, each having rather distinct origins. It is possible to regroup them
into three, because the side-jump accumulation and the anomalous distribution both originate from coordinate shifts at  scattering events; similarly,
the conventional and the intrinsic skew scatterings both appear from the asymmetry in the collision term kernel in the semiclassical Boltzmann equation.

However, such a simplified classification into only three parts was one of the reasons for a confusion. 
For example, it is customarily stated that the skew scattering always leads to 
the conductivity that depends as $1/n$ on the impurity concentration. Table 1 shows that this is not true. The intrinsic skew scattering conductivity
is independent of the impurity concentration and is parametrically very similar to the side-jump related contributions. Unjustified claims
resulted in the omission of the intrinsic skew scattering in almost all discussions of the semiclassical approach to the AHE. Such an 
omission has been repeated in the recent efforts to design the spin Hall effect theory by the analogy with previous AHE results
\cite{engel-rev}. It is also useful to distinguish the side-jump accumulation and  the anomalous distribution effects.
 The side-jump accumulation is a rather direct consequence of  coordinate shifts while the derivation of the
anomalous distribution requires several extra steps in the semiclassical theory, which were unnoticed in a few former publications.
%. For example, 
%in several former publications devoted to the side-jump, the anomalous distribution effect was not considered \cite{holstein}.  

Another confusing example from the recent terminology is the statement that the
 vertex correction, coming from a Gaussian correlated disorder in the Kubo formula 
is due to the side-jump effect only. Comparing the
 vertex correction with the semiclassical expression for the side-jump conductivity, discussed by Berger, Nozieres and others,
one would find a discrepancy because
the intrinsic skew scattering is also captured by the vertex correction and it was not considered by the older semiclassical theories that
concentrated only on the side-jump effect. 

The classification into the "intrinsic" and "extrinsic" contributions was also understood quite arbitrarily by many authors. In the present review 
we coined the word "intrinsic" for the single special contribution, which is due to unusual trajectories of wave packets in the external electric field rather than
any other mechanism that involves scatterings on impurities. This definition of the intrinsic contribution 
is justified by the fact that it follows only from the crystal band structure. Respectively, all other contributions can be called extrinsic.
%is in agreement with the original definition in the work that introduced it.
  Sometimes, in other publications \cite{nagaosa-rev} any conductivity contribution, independent of the impurity concentration $n$
is called intrinsic. This seems not a good choice of a terminology because the side-jump and the intrinsic skew scattering effects satisfy this definition but
originate from scatterings on impurities. While corresponding conductivities are independent of $n$ they can depend on other disorder parameters, for example,
impurities with different typical ranges of scattering angles can result in Hall conductivities different by a numerical factor of order unity.  
% the only extrinsic contribution then, according to the table,
%would be the conventional skew scattering that already has its name.

 There are examples where
 the "intrinsic" is associated with any effect induced by the Berry 
curvature in Bloch bands, while "extrinsic" would be due to relativistic corrections to the  impurity potential, possibly renormalized in conducting bands by
the crystal potential, as discussed by Berger, Chazalviel and others \cite{berger-1, chazalviel, nozieres}.
For example, according to this terminology all effects produced in the 2D electron system by the Rashba spin orbit coupling are intrinsic and all other effects such as due to the spin orbit part of the 
impurity potential are extrinsic. 
 In such a case all the effects discussed in this review
would be called intrinsic.
Such a terminology also seems somewhat misleading, because effects due to the disorder spin orbit coupling also can be described and classified in the same way as here when working
with the relativistic Dirac equation for electrons or, in the case of semiconductors, 
 this corresponds to working with the 8-band model without projecting all operators to the 
conducting 2-band system. Then one can work with a disorder potential free of the spin-orbit coupling and
contributions to the AHE then can be derived in the same way as in this review  \cite{chazalviel,bruno-dirac}.

Finally, there is a notion of   the "Berry phase" contribution to the AHE \cite{jungwirth-02}. Often it is identified with the intrinsic contribution as defined in this review.
However, origins of all  disorder related effects can also be traced to the Berry phase or to a nontrivial topology of Bloch bands. For example, the side jump expression and the antisymmetric 
part of the scattering rate can be expressed via a topological Pancharatnam phase \cite{sinitsyn-sj}. Therefore it is possible to speak about the Berry phase {\it mechanism} of the AHE,
which includes all the physics discussed here but to apply this terminology to a particular contribution can be misleading.

\section{Summary}

The modern semiclassical theory rigorously takes into account all known important contributions in the model of electrons in Bloch bands interacting with static impurities.
Predictions of this theory were verified with rigorous
quantum mechanical techniques. However, so far the semiclassical theory of the AHE was built to deal with electrons that
do not interact with each other. 

The state of the art is currently at the stage similar
to where the theory of electrons in metals was before Landau introduced the Fermi liquid hypothesis.
The Fermi liquid theory was originally semiclassical 
and allowed to derive many important properties of the electron state prior to systematic diagrammatic calculations.

 The important problem  now is
the effect of the nonzero Berry curvature on many-body interactions and only
recently new publications appeared that addressed it.
Thus, Haldane \cite{haldane-04} proposed that the Berry phase can be considered as the property of
quasiparticles living near the Fermi surface. 
In  \cite{shi-07} Shi {\it et al} showed the robustness of several results of the wave packet theory against $e$-$ e$ interactions.
Shindou and Balents in \cite{balents-07} studied the problem in more details by deriving the quantum kinetic equation. The interesting finding was that
the Berry curvature now acts as a pseudo-magnetic field in the extended $({\bf k},\omega)$ space.
Also, Shi and Niu \cite{shi-06} considered interacting wave packets in  bands with
the Berry curvature and found the attracting force that can induce the instability in the p-channel and thus can facilitate the unconventional superconductivity.
More generally, there is a strong similarity between properties of systems with the AHE and the p-type superconductivity and superfluidity in $^3$He-A 
\cite{mermin-80,volovik-88,volovik-97,stone-03,murakami-03}.
 The effective action  governing low-energy physics of the $p_x+ip_y$ superfluid contains a topological
contribution, leading to the Hall effect, similar to the intrinsic contribution to the AHE in metals. It should be interesting to explore the 
diffusion equation for Bogoliubov's quasiparticles on this background. For example, one
can expect to find effects similar to the side-jump and the skew scattering.  

Another poorly understood problem is the physics near the edges in systems with the AHE.
% By analogy with other types of the Hall effect one can expect that 
%the Hall conductivity is connected to the currents near the edges of the sample.
% Unlike the quantum Hall effect, the transverse conductivity in the AHE, including its
%intrinsic part, is not quantized in units $e^2/h$. This suggests that the localized edge states alone cannot be responsible for all currents near the edges in the metallic
%regime.
It is possible  that the side-jump type scatterings from the edge lead to the edge current.
 Analogous phenomena are known in geometrical optics where
the reflected beam is shifted from the incident point on the mirror \cite{goos-47,fedorov-55,imbert-77,zeldovich-92,onoda-optics,bliokh-optics,hugrass-90}.
There can be complications with this analogy because a reflection of the beam generally changes its intrinsic angular momentum \cite{hugrass-90}.
%Accumulation of such coordinate shifts of scattered wave packets from the sample edges can contribute to the total edge current, 

Recently the optically induced AHE attracted some interest both theoretically \cite{borunda-07,yao_a-07,dai-07} and experimentally \cite{exp-ahe-optic1,exp-ahe-optic2}. 
The conventional AHE needs a magnetization  in order to break the time-reversal
symmetry. This requirement can be avoided if other interactions, such as with a polarized light are introduced.
 An observation of such an optically induced Hall current would allow to 
explore the physics discussed above avoiding many difficulties in the interpretation of the standard AHE because currently ferromagnetic samples are very dirty. 
So far experimental results on this topic are controversial, with one group reported the observation of the
effect  \cite{exp-ahe-optic1} and another reported that the effect was {\it not} observed in GaAs at least up to the measurement uncertainty \cite{exp-ahe-optic2}
with results in the agreement with the classical Hanle effect.
%, while 
%a signal with the same symmetry as Hall with respect to  (i.e. Hanle effect) appears due to the micron-scale movement of the optical system by field.
The rigorous theory of the optically induced AHE is also missing, although the semiclassical theory allows to  make an insight. For example, one can expect to find an intrinsic-like contribution due to 
distorted trajectories of electrons in simultaneously applied DC and  circularly polarized AC electric fields. The absorption of a photon from a polarized light can induce side-jump-like
shifts {\it etc}. 

One more recent experiment demonstrated that the anomalous Hall conductivity can be measurable even in paramagnetic materials when electron spins
are polarized by an applied external magnetic field \cite{exp-ahe-param1}. The theoretical model of this 
effect must inevitably deal with a strong conventional Hall effect and the strict separation into conventional and anomalous Hall effects may not work anymore, for example due to 
the phase space volume correction following from wave packet equations of motion. The theory of the AHE in this regime should be upgraded.

 It is worth mentioning that new types of the AHE have been recently 
proposed for an experimental verification \cite{zeldovich-92,onoda-optics,bruno-rods, ahe-phonon} 
and the semiclassical theory can be used to describe these effects too.
 There are also suggestions of alternative mechanisms of the AHE and the spin Hall effect based on the possibility of a spin-dependent force 
\cite{hirsch-99,chudnovsky-07, shen-06}. This idea so far is based mainly on the Drude-model type of arguments and its verification by rigorous quantum 
mechanical and numerical techniques still has not been developed.  

Finally, the extra numerical and  {\em ab-initio}  study of the AHE is needed.
 So far all existing research of this kind concentrated only on the intrinsic contribution, treating discrepancies
with experiments by introducing a finite life-time of quasiparticles \cite{yao-04,yao-07}. The rigorous numerical study of the anomalous Hall conductivity in models  with a realistic disorder is still missing. In contrast,
currently there is a number of successful publications on the related spin Hall effect  \cite{murakami-she,sinova-she} that demonstrated a good agreement with existing theoretical results and allowed to extend them to
the analytically complicated and experimentally more realistic strong disorder cases \cite{nomura-05,nomura-06, moca-07, xing-07,hankiewicz-1,hankiewicz-2,nikolic-05}.
 Such numerical studies were also valuable to understand the problem of the spin accumulation near the edges. 
Similar research should be very valuable in applications to the AHE. Simple models such as the Rashba coupled 2D electron system with an out-of-plane magnetization should be under the 
reach of modern numerical algorithms.

%\begin{acknowledgments}
\ack{  I thank M. P. Anatska for the encouragement and the critical reading of this manuscript and I thank J. Sinova,  Q. Niu and A. H. MacDonald for the 
useful collaboration. 
 This work was funded in part by DOE under Contract No.\
  DE-AC52-06NA25396.  }
%\end{acknowledgments}

\end{document}